\documentclass[a4paper,11pt]{article}
\usepackage[english]{babel}
\usepackage{amsmath}
\usepackage{latexsym}

\begin{document}

\allowdisplaybreaks[1]

\title{Universality of affine formulation in General Relativity theory}
\author{Jerzy Kijowski\\
and\\
Roman Werpachowski\\
\\Center for Theoretical
Physics, Polish Academy of Sciences\\e-mail: roman@cft.edu.pl}
\date{5.I.2007}

\maketitle

\begin{abstract}
  The affine variational principle for General Relativity, proposed in
  1978 by one of us, is a good remedy for
  the nonuniversal properties of the standard, metric formulation,
  arising when the matter Lagrangian depends upon the metric
  derivatives. The affine version of the theory cures the standard
  drawback of the metric version, where the leading (second order)
  term of the field equations depends upon the matter fields and its
  causal structure violates the light cone structure of the metric.
  Choosing the affine connection (and not the metric one) as the
  gravitational configuration, simplifies considerably the canonical
  structure of the theory and is more suitable for the purposes of its
  quantization along the lines of Ashtekar and Lewandowski. We show
  how the affine formulation provides a simple method to handle
  boundary integrals in General Relativity theory.
\end{abstract}


\newcommand{\abs}[1]{\left\lvert#1\right\rvert}
\newcommand{\met}[1]{\smash{\mathring{#1}}\vphantom{#1}}
\newcommand{\mconn}{\met{\Gamma}}
\newcommand{\volfrm}{\sqrt{\abs{\det g}}}
\newcommand{\volfrmfrac}{\frac{\volfrm}{2k}}
\newcommand{\volfrmtfrac}{\frac{\volfrm}{2k}}
\newcommand{\varmat}{\text{ variation w.r. to matter field}}
\newcommand{\eqin}{\stackrel{\text{in}}{=}}
\newcommand{\metnabla}{\met{\nabla}}
\newcommand{\dens}[1]{{\mathcal{#1}}}
\newcommand{\laff}{\dens{L}_{\mathrm{A}}}
\newcommand{\lmata}{\dens{L}_\mathrm{A}^{\mathrm{mat}}}
\newcommand{\lmet}{\dens{L}_\mathrm{M}}
\newcommand{\lmatm}{\dens{L}_\mathrm{M}^{\mathrm{mat}}}
\newcommand{\lhilb}{\dens{L}_\mathrm{H}}

\newtheorem{theorem}{Theorem}

\section{Introduction}
\label{sec:introduction}

The field equations of electrodynamics may be written in a universal
form: $F=dA$ and $d{\cal F}={\cal J}$, valid also in the case of
nonlinear models. The only difference between the linear
(Maxwell) theory and its nonlinear variants (Mie, Born-Infeld,
phenomenological models of continuous media etc.) consists in the
constitutive equation, relating the electromagnetic tensor $F$
(encoding fields $E$ and $B$) with the ''induction tensor'' ${\cal
F}$ (encoding fields $D$ and $H$): the Maxwell constitutive
equation ${\cal F}= *F$ is no longer valid in the nonlinear cases.

Einstein's theory of gravity is often interpreted in a similar
way: the metric connection $\Gamma$ expressed in terms of
derivatives of the metric plays role of the homogeneous equation
$F=dA$, whereas the Einstein equations ${\cal G} = \kappa {\cal T}$
correspond to the nonhomogeneous Maxwell equation. One argues
that the Einstein tensor ${\cal G}$ is a universal differential
operator acting on the metric, whereas the matter energy-momentum
tensor ${\cal T}$ plays role of the ``source'' and depends upon
the specific matter properties, according to the ``constitutive
equations'':
\begin{equation}\label{e-m-var}
   {\cal T}^{\mu\nu}:= 2 \frac{\delta
    \lmatm}{\delta g_{\mu\nu}}
    = 2 \left( \frac{\partial
    \lmatm}{\partial g_{\mu\nu}}
    - \partial_\lambda \frac{\partial
    \lmatm}{\partial g_{\mu\nu ,\lambda}}
    \right) \ .
\end{equation}
Such an interpretation fails, however, when the matter Lagrangian
$\lmatm$ depends upon the derivatives $\partial g$ of the metric {\it
via} the Christoffel symbols, contained e.~g.~in the covariant derivatives
of the matter fields. In this case, the last term of \eqref{e-m-var}
contains the second derivatives of the metric. Hence, the Einstein
equations ${\cal G} = \kappa {\cal T}$ imply that the second order
differential operator acting on the metric is composed of two parts:
1) the universal part coming from ${\cal G}$ which, as is
generally known, respects the light-cone structure of the metric
$g$ and 2) the second order terms contained in ${\cal T}$. The
causal structure of the resulting operator depends, therefore,
upon the matter configuration and may differ considerably from the
light-cone structure of the metric. This means that the small
perturbations of a solution could, possibly, propagate outside of
the light-cone (see Section~\ref{simplified} for a more detailed
discussion of these nonuniversal features).

In the present paper we show that it is possible to restore the
universality of the theory by rewriting it in the so-called affine
formulation. It is based on the ``affine variational principle''
for General Relativity, proposed by one of us (J.K.) in 1978 \cite{new-principle}.
It consists in deriving the Einstein theory of
the combined ``gravity + matter'' system from the first order
Lagrangian density\footnote{In
  General Relativity one always deals with Lagrangian densities and not with
  scalars, since there is no volume form given {\it a priori}.
  To simplify the terminology, we will omit the word ``density''
  when discussing Lagrangians.}
\begin{equation}
  \label{eq:laff-1}
  \laff = \laff (\Gamma, \partial \Gamma ,
  \varphi, \partial \varphi) \ ,
\end{equation}
where $\Gamma$ is an arbitrary symmetric space-time connection:
$\Gamma = \left({\Gamma^\lambda}_{\mu\nu}\right)$ with
${\Gamma^\lambda}_{\mu\nu} = {\Gamma^\lambda}_{\nu\mu}$, and
$\varphi$ stands for matter field(s) of any nature. Physically,
$\Gamma$ describes the local inertial frames which are necessary
for the Einstein's interpretation of gravity in terms of the
``freely falling elevators''. Indeed, Einstein's approach to
gravity is based on the following observation: at each space-time
point $p$ the gravitational force may be eliminated and the Newton
equation for freely falling test bodies may be reduced to ${\ddot
y}^k(t)= 0$. The possibility of such an elimination is equivalent
to the existence of a ``local inertial frame'' at each point $p$
separately (global inertial frames do not exist in general).
Mathematically, such a frame may be defined as an equivalence
class of local coordinate charts. The equivalence is understood
with respect to the following relation: two charts $(y^\alpha)$
and $(x^\mu)$ are equivalent (belong to the same class) if their
mutual second derivatives $\frac {\partial^2 y^\alpha}{\partial
x^\mu \partial x^\nu}$ vanish at $p$. Gravitational field may thus
be described as a field of such local inertial frames.

For analytical purposes, every local inertial frame at $p$ may be
uniquely characterized with respect to a given generic (not
necessarily inertial!) reference chart $(x^\mu)$, by combinations
of derivatives
\begin{equation}\label{components}
    {\Gamma^\lambda}_{\mu\nu} (p):=
  \frac {\partial x^\lambda}{\partial y^\alpha} \
  \frac {\partial^2 y^\alpha}{\partial x^\mu \partial x^\nu}\ (p)
% RW: dopisa³em \partial przed x^\nu
\end{equation}
of its arbitrary representative $(y^\alpha)$. It is easy to see
that the components $\Gamma$ do not depend upon the choice of a
representative $(y^\alpha)$ of the given equivalence class which
constitutes the inertial frame in question. Moreover, they
transform like components of a \emph{symmetric} connection with respect
to changes of the chart $(x^\mu)$. The entire information about
the local inertial frame may be retrieved from the $\Gamma$'s: a
coordinate chart belongs to the privileged equivalence class at
$p$ if and only if the corresponding connection coefficients
\eqref{components} vanish at $p$. The field $\Gamma$ captures
entire information about the space-time configuration of the
gravitational field in the following sense: equation ${\ddot y}^k(t)=
0$, written in the inertial frame, is equivalent to ${\ddot
x}^\lambda + {\Gamma^\lambda}_{\mu\nu} {\dot x}^\mu{\dot x}^\nu =
0$ in an arbitrary frame. This implies that all the trajectories
of the freely gravitating test bodies may be calculated once we
know $\Gamma$. No information about the metric is necessary here.

The affine Lagrangian \eqref{eq:laff-1} contains no metric.
However, every field configuration of this theory is also equipped
with a metric tensor, carried through the momentum canonically
conjugate to $\Gamma$ which, in turn, is defined {\em via}
derivatives $\partial\laff / \partial(\partial\Gamma)$ of the
Lagrangian with respect to partial derivatives of the
configuration $\Gamma$. As will be seen in the sequel, it is
natural to use the ``covariant-density'' version of the metric
($k=8\pi G$ is the gravitational constant):
\begin{equation}\label{pi-def}
    \pi^{\mu\nu} :=\volfrmfrac  \ g^{\mu\nu} \ ,
\end{equation}
and to relate it to the derivative of the Lagrangian with respect
to a specific combination of derivatives
$\partial_\kappa{\Gamma^\lambda}_{\mu\nu}$, namely the one which
is contained in the symmetric part $K_{\mu\nu}:=R_{(\mu\nu)}=
\tfrac 12 (R_{\mu\nu} + R_{\nu\mu})$ of the Ricci tensor
$R_{\nu\mu}$ of $\Gamma$:
\begin{equation}\label{constit}
    \pi^{\mu\nu} = \frac {\partial\laff}
  {\partial K_{\mu\nu}} \ .
\end{equation}
Euler-Lagrange equations of this theory
\begin{equation}\label{eq:Eu-Lagr}
    \frac{\delta \laff}{\delta \Gamma}= 0 \ , \ \ \ \ \ \
    \frac{\delta \laff}{\delta \varphi}= 0 \ ,
\end{equation}
rewritten in terms of this metric, are equivalent to the
conventional system: ``Einstein equations + matter equations''. Of
course, only solutions with the correct signature $(-,+,+,+)$
describe physical spacetime geometries. This additional
restriction on the solutions of the above field equations must,
therefore, be imposed (see also discussion in Section
\ref{simple}, below formula \eqref{eq:kg-Ricci-2}). We stress,
however, that for many purposes, like the string theory or the
``complex heaven'' theory, different signatures or even a
complexification of Einstein theory might also be
interesting~\cite{finleyplebanski,newman}. Within this (already
classical) framework, a restriction on the signature of the metric
is thus also necessary in order to select physically admissible
solutions among all possible mathematical solutions of the theory.
As we show in the sequel, the affine formulation does not rely on
the specific signature of the metric and exists in all these
cases.

As an example take the following affine Lagrangian, depending upon
the connection, the scalar matter field $\phi$ and their
derivatives:
\begin{equation}
\label{eq:exlaff}
\laff(K, \phi, \partial \phi) = \frac{2}{m^2 \phi^2} \sqrt{\abs{\det
\left(\frac{1}{k} K_{\mu\nu} - \phi_{,\mu} \phi_{,\nu}\right)}} \ .
\end{equation}
Definition~\eqref{constit} of the canonical momenta $\pi$ implies
the following relation between the metric and the curvature:
\begin{equation}
\pi^{\mu\nu} = \frac{1}{m^2 \phi^2} \sqrt{\abs{\det \left(\frac{1}{k}
K_{\mu\nu} - \phi_{,\mu} \phi_{,\nu}\right)}} \left[ \left( \frac{1}{k}
K - \partial \phi \cdot \partial \phi \right)^{-1} \right]^{\mu\nu} \ ,
\end{equation}
where $\left( \frac{1}{k} K + \partial \phi \cdot \partial \phi
\right)^{-1}$ is the matrix inverse of $(\frac{1}{k} K_{\mu\nu} -
\phi_{,\mu} \phi_{,\nu})$. Calculating the determinant of both
sides and using definition~\eqref{pi-def}, we may obtain the following
metric tensor
\begin{equation}
  g_{\mu\nu} = \frac{2}{m^2 \phi^2} \left( \frac{1}{k} K_{\mu\nu} -
  \phi_{,\mu} \phi_{,\nu} \right) \ ,
\end{equation}
and, consequently, we have:
\begin{equation}\label{Kzg}
  K_{\mu\nu} = \frac{k}{2} m^2 \phi^2 g_{\mu\nu} + k \phi_{,\mu}
  \phi_{,\nu} \ .
\end{equation}
Einstein tensor $G_{\mu\nu} = K_{\mu\nu} - \frac{1}{2}
K_{\alpha\beta} g^{\alpha\beta} g_{\mu\nu}$ fulfils, therefore,
the following equations:
\begin{equation}\label{EKG}
  \frac 1k \   G_{\mu\nu} =  \left( \phi_{,\mu} \phi_{,\nu} - \frac{1}{2} m^2
  \phi^2 g_{\mu\nu} - g^{\alpha\beta} \phi_{,\alpha} \phi_{,\beta}
  g_{\mu\nu} \right) \ .
\end{equation}
These are Einstein equations $G = k T$ for the Klein-Gordon scalar
field, whose energy-momentum tensor $T_{\mu\nu}$ is described by
the right-hand side of the above equation. It will be proved in
the sequel (see Section \ref{simple}) that:
\begin{enumerate}
\item The first Euler-Lagrange equations in~\eqref{eq:Eu-Lagr}
implies that $\Gamma$ is the metric (Levi-Civitta) connection of $g$ (and, whence, $K_{\mu\nu} = R_{\mu\nu}$), \item The
second Euler-Lagrange equations in~\eqref{eq:Eu-Lagr} implies the
Klein-Gordon equation
\begin{equation}
\left( \Box - m^2 \right) \phi = 0 \ ,
\end{equation}
where $\Box$ is the natural d'Alembert operator for the metric
$g$.
\end{enumerate}
We conclude that \eqref{eq:exlaff} is the affine Lagrangian of the
Einstein-Klein-Gordon theory.

Even if perfectly equivalent to the conventional (metric)
formulation of the General Relativity theory, the new (affine)
formulation clarifies considerably its canonical structure. This
is due to the fact that all the boundary integrals, otherwise very
complicated and even messy, gather in a natural way to the
integral of the affine contact form ``$\pi\cdot \delta \Gamma\,$''
(analogous to ``$p \cdot\delta q$'' in canonical mechanics). This
observation simplifies enormously any manipulation of boundary
terms in General Relativity, regardless of which of the three
variational principles one begins with: 1) metric, 2) affine or 3)
metric-affine (Palatini). As an application, one obtains an
effective criterion which enables us to check which one among the
many existing quasi-local expressions for the gravitational mass
really corresponds to the field hamiltonian~
\cite{canonical-structure,Laszlo,Chrusciel}. Recently, this
technique was used to identify the first law of thermodynamics of
black holes with the energy flux due to the variation of the field
boundary data, and to generalize it to dynamical black holes~\cite{B-H}.
A similar integration of the ``$\pi \cdot \delta
\Gamma\,$'' term over a wave front gives the simplest derivation
of the dynamics of a null-like, self-gravitating shell~\cite{null-shell}.
Recently, the same techniques was used to
derive the variational principle for the ``spherically symmetric
massive shell + gravity'' system~\cite{Daniele}.

The affine point of view is confirmed by the recent developments of
the quantum gravity~\cite{qgrav-srep}. Indeed, the
fundamental idea of the Ashtekar description of the canonical
structure of the space of the Cauchy data for the Einstein equations is
that the field configuration is encoded in the connection, whereas
the metric plays role of a corresponding momentum (see
e.g.~\cite{ashtekar,ashtekar96}). This picture is much
more natural when we construct the canonical structure already
from the affine variational principle~\cite{canonical-structure}.

The affine picture in the General Relativity theory should also be taken
seriously into account when we try to quantize gravity {\em via}
path integrals (the so-called ``quantum foam'' approach).
Interpreting the scalar curvature not as a degenerate Lagrangian
(degenerate, because linear in highest derivatives!) but merely
as a standard ,,$\, p\cdot{\dot q}\,$'' term in the Legendre
transformation formula ,,$L = p\cdot{\dot q} - H$'' (see formula
\eqref{eq:aff-num}) changes completely the philosophy of path
integrals, where the integration should be performed over the space of
connections $\Gamma$ rather than over the space of metrics. This
result strongly supports the Ashtekar-Lewandowski approach to
Quantum Gravity.

Our paper is organized as follows. In Section \ref{heuristic} we
illustrate the relations between various variational principles on
a toy model of classical mechanics. In Section \ref{simple} we
briefly sketch the results of paper \cite{new-principle} and give
a simple argument for the validity of the affine picture in a
simplest case, when the matter Lagrangian does not depend upon the
derivatives of the metric, i.~e.~when the last (nonuniversal)
term in \eqref{e-m-var} vanishes. Section
\ref{sec:simpl-einst-equat} deals already with generic
Lagrangians. We construct a certain nonmetric connection $\Gamma$
and show that the Einstein equations recover their universality
when rewritten in terms of $\Gamma$. In Section
\ref{sec:constr-affine-lagr} we prove that also the matter field
equations are equivalent in both the metric and the affine
picture. Mathematically, these results are equivalent to those
proved in \cite{Torino} and \cite{ferraris1}. However, the
approach proposed here is better adapted for the purposes of the
metric theories and the proofs are new. Section
\ref{sec:conclusions} contains a brief summary of the particular
features of the affine approach. Finally, Appendix A shows how to
handle in a simple way the boundary terms in GR. The method is
based on the affine picture, but its validity is universal. One
may apply it to Lagrangian or Hamiltonian problems, whenever the
contribution from the boundary has to be taken into
account~\cite{canonical-structure,B-H,null-shell,Daniele,thermo-elastic}.
Technical proofs have been shifted to
Appendix~\ref{sec:proofs-vari-form}. We have provided an example
of the transformation between the affine and metric principles in
Appendix~\ref{sec:transf-covect}.

\section{A heuristic idea of the affine formulation}\label{heuristic}

To illustrate the relation between the metric and the affine formulation, we
consider a toy example: the one-dimensional harmonic oscillator. It is
described by two quantities: the momentum $p$ and the
configuration $q$. They fulfill the following dynamical equations:
\begin{eqnarray}
    p  &=& m \dot q \ , \label{qdot} \\
  {\dot p} &=&  - k q  \label{pdot} \ .
\end{eqnarray}
As has been noticed in \cite{gr-gauge}, the theory may be derived
from the following second order Lagrangian
\begin{equation}\label{dz}
    L_{\rm M}(p, \dot p , \ddot p ) = - \frac 1k p \ddot p - \frac 1{2k} {\dot p}^2
    - \frac 1{2m} p^2 \ ,
\end{equation}
if we keep \eqref{pdot} as a definition of an ``auxiliary
variable''
\begin{equation}
\label{eq:defq}
q := - \frac 1k  \dot p \ .
\end{equation}
Indeed, variation with
respect to $p$ reproduces the remaining equation \eqref{qdot}:
\begin{equation}\label{ww}
  \begin{split}
    \frac{\delta L_{\rm M}}{\delta p} &=\frac {{\rm d}^2}{{\rm d}t^2} \
    \frac{\partial L_{\rm M}}{\partial\ddot p} - \frac {{\rm d}}{{\rm d}t} \
    \frac{\partial L_{\rm M}}{\partial\dot p} + \frac{\partial L_{\rm M}}{\partial
      p} \\
    &= - \frac 1k \ddot p - \frac 1{m} p =\dot q - \frac 1{m} p = 0 \
    .
  \end{split}
\end{equation}
The above ``crazy'' formulation is very much analogous to the
metric formulation of relativity, where $p$ and $q$ play the roles of
the metric $g$ and the metric connection $\Gamma$ respectively.
Like the Hilbert Lagrangian in the gravity theory, the quantity $L_{\rm M}$
is linear with respect to the second derivatives and quadratic with
respect to the first derivatives of the configuration $p\,$. The
``Palatini method of variation'' consists in replacing the first
derivatives of the configuration ($\dot p$ here and $\partial g$
in the gravity theory) by ``an auxiliary variable'' ($q$ here and
$\Gamma$ in the gravity theory). This leads to the following
``Palatini version'' of (\ref{dz}):
\begin{equation}\label{Palat}
       L_{\rm P}(p, q , \dot q ) =  p \dot q - \frac k2 q^2
        - \frac 1{2m} p^2 \ .
 \end{equation}
Here, the variation with respect to $q$ reproduces its definition~\eqref{eq:defq},
whereas the variation with respect to $p$ reproduces the dynamical
equation \eqref{ww} (analogous to the Einstein equation in the
gravity theory).

Finally, using the latter equation, we may eliminate $p$
(the metric) from the Lagrangian and obtain its ``affine version''
which turns out to be the conventional Lagrangian for the
oscillator:
\begin{equation}\label{met-osc}
      L_{\rm A}( q , \dot q ) =  \frac m2 {\dot q}^2 - \frac k2 q^2\ ,
\end{equation}
numerically equal to the previous Lagrangians $L_{\rm M}$ and $L_{\rm P}$. The
above quantity $L_{\rm A}$ is as an analogue of the affine Lagrangian in
the gravity theory. Now, \eqref{pdot} is treated as the dynamical
equation, which may be derived from the Lagrangian $L_{\rm A}$, whereas
\eqref{qdot} is merely a definition of an auxiliary variable $p$.

The standard canonical structure of the theory based on $L_A$ is
fully given by the symplectic form ${\rm d}p \wedge {\rm d}
q$. Its derivation is very much analogous to the canonical
gravity derived from the affine formulation (see
\cite{canonical-structure,framework}). The derivation of
this structure from the original second order Lagrangian
\eqref{dz} or its ``Palatini version'' \eqref{Palat} is much more
difficult and leads basically to second-type constraints. We
recover the final result only when we reduce the theory with
respect to these constraints. Such a procedure is very much
analogous to the original ADM procedure~\cite{ADM} in the
gravity theory, whereas the ,,affine'' Lagrangian \eqref{met-osc}
leads directly to the correct canonical structure of the theory.

\section{A simple argument for the affine formulation of Einstein
gravity theory}\label{simple}

It was proved in~\cite{new-principle} that the Euler-Lagrange
equations~\eqref{eq:Eu-Lagr} are equivalent to the conventional system
``the Einstein equations + the matter equations'' provided the following
condition is satisfied:

\noindent {\em $\laff$ is an invariant scalar density which
depends upon a connection $\Gamma$ and its derivatives only \textit{via} the symmetric
    part $K_{\mu\nu}:=R_{(\mu\nu)}= \tfrac 12 (R_{\mu\nu} +
    R_{\nu\mu})$ of the Ricci tensor $R_{\nu\mu}$ of $\Gamma$}:
\begin{equation}\label{eq:L(K)}
    \laff = \laff (K, \varphi, \partial \varphi) \ .
\end{equation}
No metricity condition is imposed on $\Gamma$, because there is no
metric in $\laff$. This is why the Ricci tensor could possess {\em
a priori} an anti-symmetric part. However, any field configuration
of this theory carries also a metric tensor $g=(g_{\mu\nu})$,
defined in its ``contravariant-density'' version~\eqref{pi-def} by
the following equations:
\begin{equation}
  \label{eq:metric}
  \pi^{\mu\nu}  :=
  \frac{\partial \laff(K, \varphi, \partial \varphi)}{\partial
    K_{\mu\nu}} \ .
\end{equation}
The equivalence of the
two theories means that, when rewritten in terms of this metric
tensor, the field equations~(\ref{eq:Eu-Lagr}) are equivalent to the
combined ``Einstein equations + matter field equations'' for a
certain matter theory, whose matter Lagrangian is uniquely
constructed in terms of the affine Lagrangian $\laff$.

A simple argument given in \cite{new-principle} for the
equivalence of the above ``affine theory'' and the conventional
metric formulation of the General Relativity theory follows from the
observation that the gravitational part of the Euler-Lagrange
equations (\ref{eq:Eu-Lagr}) may be written as follows:
\begin{equation}\label{eq:metric-conn}
    0= - \frac{\delta \laff}{\delta \Gamma^\lambda_{\mu\nu}}=
    \partial_\kappa \left\{
    \frac{\partial \laff} {\partial K_{\alpha\beta}}
    \frac{\partial K_{\alpha\beta}} {\partial
    \Gamma^\lambda_{\mu\nu , \kappa}} \right\} -
    \frac{\partial \laff} {\partial K_{\alpha\beta}}
    \frac{\partial K_{\alpha\beta}} {\partial
    \Gamma^\lambda_{\mu\nu}}
    \ ,
\end{equation}
where we denote
\[
\Gamma^\lambda_{\mu\nu , \kappa} :=
\partial_\kappa\Gamma^\lambda_{\mu\nu} \ .
\]
The definition of the Ricci tensor
\begin{equation}\label{eq:K}
    K_{\mu\nu} := {\Gamma^\lambda}_{\mu\nu,\lambda} -
    {\Gamma^\lambda}_{(\mu|\lambda |, \nu)} +
    {\Gamma^\lambda}_{\alpha\lambda} {\Gamma^\alpha}_{\mu\nu}
    -{\Gamma^\lambda}_{\alpha\nu} {\Gamma^\alpha}_{\mu\lambda}
\end{equation}
implies that $K$ is linear in
$\Gamma^\lambda_{\mu\nu , \kappa}$ and quadratic in
$\Gamma^\lambda_{\mu\nu}$. Hence, we end up with an equation of the
type
\begin{equation}\label{E-L-equ}
    \partial \pi - \Gamma \cdot \pi = 0 \ .
\end{equation}
As will be seen in the sequel (see
Appendix~\ref{sec:affine-principle}), the term ``$\Gamma \cdot
\pi$'' in (\ref{E-L-equ}) is precisely what we need to convert the
first term ``$\partial \pi$'' into the covariant derivative
$\nabla \pi$ with respect to $\Gamma$. Finally, the gravitational
part \eqref{eq:metric-conn} of the Euler-Lagrange equations
(\ref{eq:Eu-Lagr}) reads: $\nabla_\lambda \pi^{\mu\nu} = 0$ or,
equivalently,
\begin{equation}\label{eq:metricity-1}
    \nabla_\lambda g_{\mu\nu} = 0 \ ,
\end{equation}
which implies that $\Gamma$ must be equal to the Levi-Civita
(metric) connection of $g$. Metricity condition for the connection
(not postulated {\em a priori}\ ) is, therefore, derived from the
Lagrangian as the field dynamics (cf.~\eqref{pdot} in our toy
model). On the other hand, Einstein equations, treated in the
conventional approach as field dynamics, are derived here {\em a
priori} from the Lagrangian as the constitutive
relation~(\ref{eq:metric}) between the matter fields and the Ricci
curvature $R_{\mu\nu}$ (which, due to the metricity condition
(\ref{eq:metricity-1}), reduces to $K_{\mu\nu}$) --
cf.~\eqref{qdot} in the toy model.

To analyze also the matter dynamics $\tfrac{\delta \laff}{\delta
\varphi}= 0$, it is convenient to perform the following Legendre
transformation:
\begin{equation}
  \label{eq:lmata-1}
  \lmata (\pi,\varphi,\partial \varphi):= \laff
  (K,\varphi,\partial \varphi) - \pi^{\mu\nu}
  K_{\mu\nu} \ ,
\end{equation}
where $K=K(\pi,\varphi,\partial \varphi)$ on the right-hand side
are just the ``Einstein equations'', obtained by solving
algebraically equations (\ref{eq:metric}) with respect to $K$
(similarly as it was done in \eqref{Kzg} for the
Einstein-Klein-Gordon case). It is a standard property of the
Legendre transformation between mutually conjugate objects
$K_{\mu\nu}$ and $\pi^{\mu\nu}$, that equation $\pi =
\partial \laff / \partial K$ is converted\footnote{In classical mechanics
equation $p=\partial L / \partial {\dot q}$ converts to equation
$-{\dot q} = \partial (-H) / \partial p$, where $-H=L-p{\dot q}$.}
into equation $K =-
\partial \lmata / \partial \pi$. Hence, equations~(\ref{eq:metric})
are equivalent to
\begin{equation}
  \label{eq:Einst-lmata}
  K_{\mu\nu} = - \frac{\partial \lmata}{\partial \pi^{\mu\nu}} \ .
\end{equation}
Moreover, the equivalence
\begin{equation}\label{eq:equival1}
    \left( \frac{\delta \laff}{\delta \varphi}= 0
    \right) \Longleftrightarrow
    \left( \frac{\delta \lmata}{\delta \varphi}= 0
    \right)
\end{equation}
may be easily proved (see \cite{new-principle} or
Appendix~\ref{sec:affine-principle}). This means that $\lmata$
obtained from (\ref{eq:lmata-1}) plays role of the matter
Lagrangian for the field $\varphi$. According to
(\ref{eq:lmata-1}) the numerical value of the affine Lagrangian
 equals
\begin{equation}\label{eq:aff-num}
    \laff = \lmata + \pi^{\mu\nu} K_{\mu\nu} =
    \lmata + \volfrmfrac g^{\mu\nu} R_{\mu\nu}  \ ,
\end{equation}
and, whence, coincides with the conventional Hilbert Lagrangian, with
the metric tensor eliminated by the means of relation~(\ref{eq:metric}).

The following examples of the affine Lagrangians were
analyzed in detail:
\begin{enumerate}
\item A general nonlinear Einstein-Klein-Gordon theory~\cite{new-principle}:
\begin{equation}\label{eq:E-KG}
    \laff(K,\phi,\partial\phi) = \frac{1}{U(\phi)} \sqrt{
    | \det (\tfrac 1k K_{\mu\nu} -  \phi_{,\mu} \phi_{,\nu} ) |}  \ ,
\end{equation}
where $\phi_{,\mu} =\partial_\mu \phi$ and $U(\phi)$ is a scalar
function. In particular $U(\phi) = \frac{1}{2} m^2 \phi^2$ for the
linear Klein-Gordon theory, cf.~\eqref{eq:exlaff}.
%\cite{new-principle}). % or Appendix~\ref{sec:affine-vers-einst}

\item The Einstein-Maxwell theory~\cite{gr-gauge}
\begin{equation}\label{eq:E-M}
    \laff(K,\partial A) = - \frac 14 \sqrt{|\det K_{\rho\sigma}|}
     \ K^{\mu\nu}
    K^{\alpha\beta} F_{\mu\alpha} F_{\nu\beta} \ ,
\end{equation}
where $A_\mu$ is the electromagnetic potential of the
electromagnetic field $F_{\mu\nu}=\partial_\mu A_{\nu} -
\partial_\nu A_{\mu}$, whereas $K^{\mu\nu}$ is the
contravariant tensor, inverse to $K_{\alpha\beta}$ (i.~e.~defined
by: $K^{\mu\alpha}K_{\alpha\beta}= \delta^\mu_\beta$).

\item The vacuum Einstein theory:
\begin{equation}\label{eq:vacuum}
    \laff(K) = - \frac {1}{k \Lambda}
    \sqrt{|\det K|} \ ,
\end{equation}
where $\Lambda$ is the cosmological constant~\cite{gr-gauge}. % or Appendix~\ref{sec:affine-lagr-vacu}

\item The Einstein-Euler equations describing a self-gravitating
    barotropic fluid:
    \begin{equation}
      \label{eq:barotropic}
      \laff(K,\xi ,\partial\xi)=- \sqrt{|\det K|}
      \ f\left( \frac {K_{\mu\nu} j^\mu j^\nu}{\det K} \right) \ ,
    \end{equation}
where the function $f$ uniquely implies the constitutive equation
of the fluid~\cite{kijowski79} and $j$ is the vector density
representing the matter current of the fluid. The latter is
uniquely defined by the fluid configuration $\xi$ and its partial
derivatives, without any knowledge of the metric (see
\cite{thermo-elastic}). In particular, the linear function $f(y)=
\tfrac k8 y$ corresponds to the case of a dust (whose pressure
vanishes identically: $p\equiv 0$).
\end{enumerate}

The following observation concerning these models seems to hold
universally: ``what is simple in the metric picture becomes, in
general, complicated in the affine picture and {\em vice versa}''.
This may be a good leading principle in the search for
fundamental, unified theories of various interactions. Indeed, the
simplicity of the Lagrangian is an important feature of
fundamental theories. For the gravitational interaction, affine
approach looks much more fundamental than the metric approach.

Each of the affine theories \eqref{eq:E-KG}--\eqref{eq:barotropic}
is characterized by the fact that, in the
conventional metric formulation, its matter Lagrangian $\lmatm$
does not contain the derivatives of the metric tensor because no
``covariantization'' of the partial derivatives of the matter
field is necessary:
\begin{equation}\label{eq:part-fi}
    \lmatm = \lmatm (g,\varphi, \partial \varphi) \ .
\end{equation}
To derive the affine formulation for the matter fields of this type, it is
useful to rewrite the conventional Einstein equations in a simpler form,
analogous to (\ref{eq:Einst-lmata})\footnote{By putting a circle ``\
  $\mathring{}$\ '' over a geometric object, we indicate its
  metricity. Hence, by $\met{\nabla}$ we mean the covariant derivative
  taken with respect to the metric connection $\mconn$,
  ${\mconn^\lambda}_{\mu\nu} \equiv \lbrace {}^\lambda_{\mu\nu}
  \rbrace$.}:
\begin{equation}
  \label{Einst1}
  \left( \met{K}_{\mu\nu} = \right)
  \met{R}_{\mu\nu} = - \frac{\partial \lmatm
  }{\partial \pi^{\mu\nu}} \ .
\end{equation}
Indeed, let us take the standard metric Lagrangian $\lmet= \lhilb +
\lmatm$, where $\lhilb$ is the Hilbert Lagrangian
\begin{equation}
  \label{eq:Hilbert-Lagrangian}
  \lhilb := \volfrmfrac \met{R} = \pi^{\mu\nu} \met{K}_{\mu\nu} \ .
\end{equation}
The Einstein equations are derived as the Euler-Lagrange equations for
$\lmet$:
\begin{equation}
  \label{eq:EL-metric1}
  0 = \frac{\delta \lmet}{\delta g_{\mu\nu}} = - \frac{1}{2k}
  \met{\dens{G}}^{\mu\nu} + \frac{\partial
    \lmatm(g,\varphi,\partial\varphi)}{\partial g_{\mu\nu}} \ ,
\end{equation}
because the variation of the Hilbert Lagrangian is proportional to the
Einstein tensor density
\[
\met{\dens{G}}^{\mu\nu}:=\sqrt{|\det g
|}\left( \met{R}^{\mu\nu} - \tfrac 12 \met{R} g^{\mu\nu}\right) \ .
\]
After substituting the variable $\pi$ instead of $g$, we obtain:
\begin{equation}
  \label{eq:ELmetric1-pi}
  \begin{split}
    \met{\dens{G}}^{\mu\nu} &=  2k \frac{\partial
      \pi^{\rho\sigma}}{\partial g_{\mu\nu}} \frac{\partial \lmatm
      (\pi,\varphi,\partial\varphi)}{\partial \pi^{\rho\sigma}} \\
    &= \sqrt{|\det g|}
    \Bigl( g^{\mu\rho} g^{\nu\sigma} - \tfrac{1}{2} g^{\mu\nu}
    g^{\rho\sigma} \Bigr) \frac{\partial
      \lmatm(\pi,\varphi,\partial\varphi)}{\partial \pi^{\rho\sigma}}
    \ ,
\end{split}
\end{equation}
which, after lowering the free indices, subtracting half of the trace
and dividing by $\sqrt{|\det g|}$, gives precisely~(\ref{Einst1}).
By identifying the matter Lagrangians in both formulations:
\[
\lmatm(g,\varphi,\partial\varphi) =
\lmata(g,\varphi,\partial\varphi) \ ,
\] we get~(\ref{eq:Einst-lmata}),
which proves that the theory may now be easily converted to the
affine picture {\em via} the inverse Legendre transformation~(\ref{eq:aff-num}),
where $g = g(K,\varphi, \partial \varphi)$
must be calculated as an implicit function from~(\ref{Einst1}).

%% Klein-Gordon

To illustrate the above issues, consider again the
Einstein-Klein-Gordon theory. Its generalized (nonlinear) version  is
defined by the following metric matter Lagrangian:
\begin{eqnarray}
  \lmatm &=& \sqrt{|\det g|} \left( - \frac{1}{2} g^{\mu\nu} \phi_{,\mu}
    \phi_{,\nu} - U(\phi) \right) \nonumber\\
    \label{eq:klein-gordon-lmatm}
      &=& - k \pi^{\mu\nu} \phi_{,\mu}
    \phi_{,\nu} - 4 k^2 \sqrt{|\det \pi|} \ U (\phi)\ ,
\end{eqnarray}
where $U(\phi)$ is an arbitrary positive, convex function of
$\phi$. In particular, $U(\phi) = \frac 12 m^2\phi^2$ leads to the
linear theory, governed by equation \eqref{EKG}. We have used the
formula $|\det g| = 16 k^4 |\det \pi|$, which is an immediate
consequence of the definition \eqref{pi-def} of $\pi$. According
to formula~\eqref{Einst1} the Einstein equations may be written in
the following form:
\begin{equation}
  \label{eq:kg-Ricci}
  K_{\mu\nu} = - \frac{\partial \lmatm
  }{\partial \pi^{\mu\nu}}= k \phi_{,\mu} \phi_{,\nu} + 2k^2 U \sqrt{|\det
    \pi|} \left( \pi^{-1} \right)_{\mu\nu}  \ .
\end{equation}
It enables us to calculate the metric tensor (the momentum canonically
conjugate to $\Gamma$) in terms of ,,velocities'' ($K$ and
$\partial \phi$):
\begin{equation}
  \label{eq:kg-Ricci-2}
   g_{\mu\nu}=2 k \sqrt{|\det \pi|} \left( \pi^{-1} \right)_{\mu\nu} =
   \tfrac 1U \left( \tfrac 1k K_{\mu\nu} -  \phi_{,\mu} \phi_{,\nu}
   \right)  \ .
\end{equation}
We see that the tensor $\tfrac 1k K_{\mu\nu} -  \phi_{,\mu}
\phi_{,\nu} $ is proportional to the metric tensor. The physical
signature requirement for $g$ implies that only those
configurations $(\phi ,\Gamma )$ for which $\left(\tfrac 1k
K_{\mu\nu} - \phi_{,\mu} \phi_{,\nu} \right)$ has the signature
$(-,+,+,+)$ have to be taken into account.

To calculate the total affine Lagrangian of the theory, we perform
the Legendre transformation:
\begin{equation}
  \label{eq:kg-laff}
  \begin{split}
    \laff &= \lmata + K_{\mu\nu} \pi^{\mu\nu} = \lmata + k
    \pi^{\mu\nu} \phi_{,\mu} \phi_{,\nu} + 8 k^2 \sqrt{|\det \pi|}\
    U(\phi) \\
    &= 4 k^2 \sqrt{|\det \pi|}\  U(\phi) \ .
  \end{split}
\end{equation}
Calculating the determinant of both sides of~\eqref{eq:kg-Ricci-2}
gives the relation
\begin{equation}
  \label{eq:kg-detpi-Ricci}
  4 k^2 U^2 \sqrt{| \det \pi |} =  \sqrt{
    | \det ( \tfrac 1k K_{\mu\nu} -  \phi_{,\mu} \phi_{,\nu} ) |} \ .
\end{equation}
Inserting it into~(\ref{eq:kg-laff}) gives~(\ref{eq:E-KG}).

%% koniec Kleina-Gordona

Unfortunately, this procedure does not work for general matter
fields, where the matter Lagrangian contains covariant
derivatives:
\begin{equation}\label{eq:cov-fi}
    \lmatm = \lmatm (g,\varphi, \met\nabla \varphi)
    = \lmatm (g,\mconn,\varphi, \partial \varphi) \ .
\end{equation}
As an example we may consider the vector field $\varphi =
(\varphi^\mu)$, which needs covariant derivatives $(\met\nabla_\nu
\varphi^\mu= \partial_\nu \varphi^\mu +
{\met\Gamma}^\mu_{\nu\lambda}\varphi^\lambda)$ to construct an
invariant quantity depending on its first derivatives.

On the other hand, a general affine Lagrangian
\begin{equation}\label{eq:aff-general}
    \laff = \laff(K,\varphi,\nabla \varphi) =
    \laff(K,\Gamma,\varphi,\partial \varphi) \ ,
\end{equation}
implies that the connection $\Gamma$ differs from the metric
connection $\mconn$, due to the nonhomogeneous term
\begin{equation}
  \label{eq:Jdef}
  {\dens{J}_\lambda}^{\mu\nu} := \frac{\partial \laff (K, \Gamma,
    \varphi, \partial \varphi )}{\partial {\Gamma^\lambda}_{\mu\nu}} \ ,
\end{equation}
which arises on the right-hand side of~(\ref{eq:metric-conn}). In
fact, the gravitational part of the field equations~(\ref{eq:Eu-Lagr})
may now be rewritten (see~\cite{new-principle} or
Appendix~\ref{sec:affine-principle}) as follows:
\begin{equation}
  \label{eq:ELGamma}
  \nabla_\lambda \pi^{\mu\nu} = {\dens{J}_\lambda}^{\mu\nu} -
  \tfrac{2}{3} \delta^{(\mu}_\lambda {\dens{J}_\rho}^{\nu)\rho} \ ,
\end{equation}
instead of~(\ref{eq:metricity-1}). For a nonvanishing derivative~(\ref{eq:Jdef}),
this implies $\Gamma \ne \mconn $. Hence, the metric and
the affine theories look quite different in a generic case.

An (apparently paradoxical) result was proved in~\cite{Torino}:
the metric and the affine formulations of General Relativity
Theory are always equivalent! More precisely: a generic affine
theory~(\ref{eq:aff-general}), when rewritten in terms of the
metric and the metric connection $\mconn$, is equivalent to the
conventional Einstein metric theory. This implies an implicit
equation which, when solvable, enables us to convert a ``minimally
coupled to gravity'' matter theory~(\ref{eq:cov-fi}) into an
affine theory for a certain nonmetric connection $\Gamma$. In
both cases the nonmetricity tensor
\[
f:=\Gamma - \mconn
\]
is uniquely defined by the matter fields.

In the present paper we proceed in the opposite direction: we show
that the (very troublesome) second order derivatives of the
metric, contained in the conventional energy-momentum tensor of
the matter, together with the metric Ricci tensor, sum up in a
natural way to the (nonmetric) Ricci tensor of a new connection
$\Gamma$. This way we obtain a much simpler proof of the above
equivalence theorem. But, above all, this result provides a simple
physical interpretation of the nonmetric connection. As will be
seen in the next section, the transition between the metric and
the affine picture consists in gathering the second order
derivatives appearing in the Einstein equations (those contained
in the metric Ricci tensor and those contained in the matter
energy-momentum tensor \eqref{e-m-var}) into a single object: the
nonmetric Ricci tensor.

The proof will be carried out in two steps: first, starting with
a generic matter Lagrangian $\lmatm$ of the conventional metric
theory, we show how to simplify the Einstein equations by writing them
in the universal form~(\ref{eq:Einst-lmata}) for a certain
``affine matter Lagrangian'' $\lmata$, in general different from
$\lmatm$; second, we prove that the functional $\lmata$
constructed in the first step generates the same equations for
the matter fields and, therefore, is the correct affine matter
Lagrangian. Thus, the Legendre transformation~\eqref{eq:aff-num}
leads directly to the affine formulation of the same theory.
Detailed calculations (with examples) are also given in
\cite{werpachowski}.

\section{The simplified form of the Einstein equations}\label{simplified}
\label{sec:simpl-einst-equat}

In this section we analyze in detail the nongeneric properties of
the Einstein equations in the standard formulation and propose a
remedy for them. We begin with a conventional, metric Lagrangian
\begin{equation}\label{generic-metric}
    \lmet =\lhilb + \lmatm \ ,
\end{equation}
where $\lhilb$ is the Hilbert Lagrangian given by
(\ref{eq:Hilbert-Lagrangian}) and the matter Lagrangian
\[
  \lmatm=\lmatm
  (g,\mconn,\varphi, \partial \varphi) =\lmatm(g,\partial
  g,\varphi,\partial\varphi)
\]
does not belong to the special class~(\ref{eq:part-fi}), which was
necessary to apply directly the affine theory. The variation of
$\lhilb$ with respect to $g_{\mu\nu}$ produces $-\tfrac
1{2k}\met{\dens{G}}^{\mu\nu}$ and, therefore, instead of
\eqref{eq:EL-metric1}, we obtain the following Einstein equations
(again, we denote $\partial_\lambda
g_{\mu\nu}=g_{\mu\nu,\lambda}$):
\begin{equation}
  \label{eq:Einstein-equations-2}
  \frac 1{k} \met{\dens{G}}^{\mu\nu} = 2 \left( \frac{\partial
    \lmatm(g,\partial g, \varphi, \partial\varphi)}{\partial g_{\mu\nu}}
    - \partial_\lambda \frac{\partial
    \lmatm(g,\partial g, \varphi, \partial\varphi)}
    {\partial g_{\mu\nu,\lambda}} \right) \ ,
\end{equation}
where the quantity on the right-hand side is called ``the total
metric energy-momentum tensor'' of the matter. The above form of
the Einstein equations displays a weak point of the theory in its
metric formulation: the lack of universality. Indeed, we cannot
treat the left-hand side as the universal second order
differential operator acting on the metric field $g$ and the right
hand side as the ``source'' carried by the matter, because the last
term above also contains, in general, the second order derivatives of
the metric. This means that, even qualitatively, the properties of the
above Einstein equations are not universal and depend heavily on
the properties of the matter. In particular, the common belief
that the weak perturbations of a solution obey its light cone
structure is, in general, false because the second order
derivatives of the metric, coming from the right-hand side, may
change completely the causal structure carried by the Einstein
tensor on the left-hand side.

As will be shown later, the affine formulation eliminates this
phenomenon: second order operators acting on the configuration
fields $\Gamma$ and $\varphi$ are universal and the interaction
between matter and gravity is described by the sources which are
of the first differential order. We are going to rewrite the Einstein
equations in a form well adapted for this purpose.

Another bad feature of equations~(\ref{eq:Einstein-equations-2})
is that the splitting of the right-hand side into the two terms is
not coordinate invariant, but an immediate remedy for this is the standard
Palatini trick~\cite{palatini}: we take $g$ and $\mconn$ as
independent variables, instead of $g$ and $\partial g$. Defining
\begin{equation}
  \label{eq:metJ-definition}
  {\met{\dens{J}}_\lambda}^{\mu\nu} := \frac{
    \partial\lmatm(g,\mconn,\varphi,\partial\varphi)}
    { \partial {\mconn^\lambda}_{\mu\nu}}
\end{equation}
we may rewrite the Einstein equations (\ref{eq:Einstein-equations-2})
in the following, covariant form:
\begin{equation}
  \label{eq:Einst3}
  \frac 1k \met{\dens{G}}^{\mu\nu} =
  2  \frac{\partial
  \lmatm(g,\mconn,\varphi,\partial\varphi)}{\partial g_{\mu\nu}} -
   \metnabla_\beta \left( 2 g^{\alpha(\mu}
    {\met{\dens{J}}_\alpha}^{\nu)\beta} - g^{\alpha\beta}
    {\met{\dens{J}}_\alpha}^{\mu\nu} \right) \ .
\end{equation}
A simple, algebraic proof of the equivalence between
(\ref{eq:Einstein-equations-2}) and~(\ref{eq:Einst3}) is contained
in Appendix~\ref{proof}. Further simplification of this equations
is obtained if we encode the information about the metric in the
density $\pi$, instead of $g$. Using the same arguments as in
formula~\eqref{eq:ELmetric1-pi}, we obtain the following,
equivalent form of the Einstein equations:
\begin{equation}
  \label{eq:Einst5}
  \begin{split}
  &\met{K}_{\mu\nu} \\
  &= - \frac{\partial \lmatm ( \pi, \mconn,\varphi,\partial\varphi )}{\partial
    \pi^{\mu\nu}} - k\frac{g_{\mu\rho} g_{\nu\sigma} - \tfrac{1}{2}
    g_{\mu\nu} g_{\rho\sigma}}{\sqrt{|\det g|}}
\metnabla_\beta
    \left( 2 g^{\alpha(\rho}
      {\met{\dens{J}}_\alpha}^{\sigma)\beta} - g^{\alpha\beta}
      {\met{\dens{J}}_\alpha}^{\rho\sigma} \right) \, .
  \end{split}
\end{equation}
If $\lmatm$ does not depend on $\mconn$, we have
${\met{\dens{J}}}=0$ and equations~(\ref{eq:Einst5}) reduce to
(\ref{eq:Einst-lmata}). In this case, the affine and metric
principles are equivalent, provided we identify $\lmatm$ with
$\lmata$, $\mconn$ with $\Gamma$ and $\met{K}$ with $K$. However,
the status of the metricity condition $\nabla g \equiv 0$ changes
while passing from $\lmatm$ to $\lmata$. In the metric picture
this condition is assumed {\it a   priori}, but in the affine
picture it is derived from the variational principle as the field
equations for an \emph{independent} variable $\Gamma$ (cf. the
status of equation~\eqref{pdot} in the ``metric'' and the
``affine'' formulation of the oscillator theory).

Consider now a generic case $\met{\dens{J}}\ne 0$. No direct
identification of the metric and affine variables is possible.
However, we are going to incorporate the last ``$\metnabla
\met{\dens{J}}$'' term from the right side of~(\ref{eq:Einst5}),
together with $\met{K}$ on the left side, into the symmetric part
$K$ of the Ricci tensor of a certain nonmetric connection
$\Gamma$, which is defined in the sequel. In this way, the conventional
Einstein equations~(\ref{eq:Einst5}) will be rewritten in a
universal form, analogous to~(\ref{eq:Einst-lmata}):
\begin{equation}
  \label{eq:Einst-aff}
  K_{\mu\nu} = - \frac{\partial
  \lmata(\pi,\Gamma,\varphi,\partial\varphi)}
  {\partial \pi^{\mu\nu}} \ ,
\end{equation}
where the ``affine matter Lagrangian'' $\lmata$ will be
specified in the sequel too.

To prove that such a simplification is possible, consider any
symmetric connection $\Gamma$ and its nonmetricity tensor $f$
defined as follows:
\begin{equation}
  \label{eq:nonmetricity-tensor}
  {f^\lambda}_{\mu\nu} := {\Gamma^\lambda}_{\mu\nu} - {\mconn^\lambda}_{\mu\nu} \ .
\end{equation}
It is a matter of simple calculations (see
Appendix~\ref{sec:proof-formula}) that the symmetric part $K$ of
the affine Ricci tensor of $\Gamma$ can be expressed as a sum of
the metric Ricci tensor $\met{K}$ and the following combination of
$f$ and $\met\nabla f$:
\begin{equation}
  \label{eq:rozklad-Ricci}
    K_{\mu\nu} =
    \met{K}_{\mu\nu}
    + \met\nabla_\rho {f^\rho_{\mu\nu} - \met\nabla_{(\mu}}
    f^\rho_{\nu)\rho} +
    {f^\rho}_{\sigma\rho}{f^\sigma}_{\mu\nu} -
    {f^\rho}_{\sigma\mu}{f^\sigma}_{\rho\nu} \ .
\end{equation}
It is easy to check that the differential terms ``$\met\nabla f$''
in the above equations match perfectly the ``$\metnabla
\met{\dens{J}}$'' terms from the right side of (\ref{eq:Einst5})
if (and only if) we choose the nonmetricity $f$ as the following,
linear function of $\met{\dens{J}}$:
\begin{equation}
  \label{eq:f(J,g)}
\begin{split}
  {f^\lambda}_{\mu\nu} &:= \frac{k}{\volfrm} \bigl[
  2{\met{\dens{J}}_{(\mu}}^{\rho\lambda}g_{\nu)\rho} -
  {\met{\dens{J}}_\rho}^{\rho\lambda}g_{\mu\nu} -
  {\met{\dens{J}}_\rho}^{\sigma\tau}g^{\rho\lambda} \left(
    g_{\mu\sigma}g_{\nu\tau} - \tfrac{1}{2}g_{\sigma\tau}g_{\mu\nu}
  \right) \\ &\ + \tfrac{2}{3}
  \delta^\lambda_{(\mu}{\met{\dens{J}}_\rho}^{\rho\sigma}g_{\nu)\sigma}
  - \delta^\lambda_{(\mu}
  {\met{\dens{J}}_{\nu)}}^{\rho\sigma}g_{\rho\sigma} \bigr] \ .
\end{split}
\end{equation}
To be able to convert (according to (\ref{eq:rozklad-Ricci}))
$\met{K}$ on the left-hand side of (\ref{eq:Einst5}) into $K$ and
thus to rewrite the Einstein equations in the universal form
(\ref{eq:Einst-aff}), we need the algebraic terms ``$f \cdot f$'' in
\eqref{eq:rozklad-Ricci}, which are still missing. We are going to prove
that they may be easily obtained if we replace the ``metric matter
Lagrangian'' $\lmatm$ by the following ``affine matter
Lagrangian'' $\lmata$:
\begin{equation}
  \label{eq:lmata-def}
  \lmata(\pi,\Gamma,\varphi,\partial\varphi) :=
  \lmatm - \pi^{\mu\nu} \left( {f^\rho}_{\sigma\rho}
    {f^\sigma}_{\mu\nu} -
    {f^\rho}_{\sigma\mu}{f^\sigma}_{\rho\nu} \right) \ .
\end{equation}
This definition has to be understood as follows. We treat~(\ref{eq:nonmetricity-tensor})
as the implicit equations for
$\Gamma$ in terms of $\mconn$ and the remaining quantities
% RW: zamieni³em miejscami \mconn i \Gamma w powy¿szym wierszu (teraz uwa¿am ¿e jest dobrze)
($\pi,\varphi$ and $\partial\varphi$):
\begin{equation}
    \label{eq:Gamma=mconn+f}
    \Gamma = \mconn + f(\pi,\mconn,\varphi,\partial\varphi) \ ,
\end{equation}
where $f=f(\pi,\mconn,\varphi,\partial\varphi)$ is defined by
(\ref{eq:f(J,g)}) and (\ref{eq:metJ-definition}). This yields us
an implicit relation
$\mconn=\mconn(\pi,\Gamma,\varphi,\partial\varphi)$. Using this
relation we eliminate $\mconn$ from $\lmatm$ and $f$ on the right
side of (\ref{eq:lmata-def}) and, finally, calculate
$\lmata$ as a function of $\pi,\Gamma,\varphi$ and $\partial\varphi$.

The following theorem shows that the above procedure leads,
indeed, to the universal form (\ref{eq:Einst-aff}) of the Einstein
equations:
\begin{theorem}\label{simpl}
  The Einstein equations~(\ref{eq:Einstein-equations-2}) (or, equivalently,
  (\ref{eq:Einst5})) generated by the metric matter Lagrangian
  $\lmatm$ are equivalent\footnote{In fact, the validity of this result goes
  far beyond the generic case, when all the implicit
  functions considered above are well defined. The nongeneric case may
  also be treated in terms of symplectic relations~\cite{framework}.
  The only difference is that additional Lagrangian constraints may arise
  in the nongeneric case.} to the equations
  \[
  K_{\mu\nu} = - \frac{\partial \lmata(\pi,\Gamma,\varphi,\partial\varphi)}
  {\partial \pi^{\mu\nu}} \ ,
  \]
  for the symmetric part $K$ of the Ricci tensor of a (possibly nonmetric)
  connection $\Gamma = \mconn + f$, where the nonmetricity $f$ is
  uniquely defined by the metric and the matter fields $\varphi$ {\em via}
  equations~(\ref{eq:f(J,g)}) and~(\ref{eq:metJ-definition}).
  The affine matter Lagrangian $\lmata$ generating this equations is defined
  by equation~(\ref{eq:lmata-def}).
\end{theorem}
The above statement means that for any solution of the Einstein equations~(\ref{eq:Einstein-equations-2}),
the corresponding (possibly
nonmetric) connection $\Gamma$ (defined uniquely by equations
(\ref{eq:Gamma=mconn+f}),~(\ref{eq:f(J,g)}) and~(\ref{eq:metJ-definition}))
satisfies the universal, simplified
form~(\ref{eq:Einst-aff}) of the Einstein equations. The opposite is also
true: for any solution of equations~(\ref{eq:Einst-aff}) generated by
a given affine matter Lagrangian $\lmata$, the metric
connection satisfies the Einstein equations in the conventional form
(\ref{eq:Einstein-equations-2}) generated by the corresponding metric
matter Lagrangian $\lmatm$ and related with the former by
equation~(\ref{eq:lmata-def}). We shall see in the sequel that also
the matter equations for $\varphi$, generated respectively by $\lmatm$
and $\lmata$, are equivalent.

To conclude this Section we prove
Theorem~\ref{simpl}. For this purpose we shall
write the field equations in terms of the derivatives of
$\lmatm(\pi,\mconn,\varphi,\partial\varphi)$ with respect to its
arguments. Indeed, we already have
\begin{equation}
  \begin{split}
  \frac{\partial \lmatm }{\partial\pi^{\mu\nu}} &=
  -  \met{K}_{\mu\nu}
  - k\frac{g_{\mu\rho}
      g_{\nu\sigma} - \tfrac{1}{2} g_{\mu\nu}
      g_{\rho\sigma}}{\sqrt{|\det g|}}
\metnabla_\beta \left( 2
      g^{\alpha(\rho} {\met{\dens{J}}_\alpha}^{\sigma)\beta} -
      g^{\alpha\beta} {\met{\dens{J}}_\alpha}^{\rho\sigma} \right) \\
   &= - \left( \met{K}_{\mu\nu}
    + \met\nabla_\rho {f^\rho_{\mu\nu} - \met\nabla_{(\mu}}
    f^\rho_{\nu)\rho} \right) \ , \label{deriv-pi}
  \end{split}
\end{equation}
due to equations~(\ref{eq:Einst5}) and the definition
(\ref{eq:f(J,g)}) of $f$. Also, $\met{\dens{J}}$ is equal to
$\partial\lmatm/\partial\mconn$, due to equations
(\ref{eq:metJ-definition}). Moreover, the Euler-Lagrange equations for
the matter fields may be written as
${p^\kappa}_{,\kappa}=\partial\lmatm/\partial\varphi$, where
\[
p^\kappa := \partial\lmatm/\partial\varphi_{,\kappa}
\]
is the momentum canonically conjugate to the matter fields $\varphi$
and, as usual,
\[
\varphi_{,\kappa}:= \partial_\kappa \varphi \ ,
\]
and
\[
{p^\kappa}_{,\kappa}:= \partial_\kappa{p^\kappa} \ .
\]
These four field equations (at each space-time point $x$) may be written in
terms of a single equation imposed on the total differential of
$\lmatm$, treated as a function of the field variables
$\pi^{\mu\nu},{\mconn^\lambda}_{\mu\nu},\varphi$ and $\partial_{\kappa}\varphi$:
\begin{equation}
  \label{eq:var-lmatm}
    {\rm d} \lmatm
    = - \left( \met{K}_{\mu\nu}
    + \met\nabla_\rho {f^\rho_{\mu\nu} - \met\nabla_{(\mu}}
    f^\rho_{\nu)\rho} \right)
       {\rm d} \pi^{\mu\nu}
    + {\met{\dens{J}}_\lambda}^{\mu\nu} {\rm d} {\mconn^\lambda}_{\mu\nu} +
    p^\kappa {\rm d} \varphi_{,\kappa} + {p^\kappa}_{,\kappa} {\rm d}
    \varphi \ .
\end{equation}
Using (\ref{eq:nonmetricity-tensor}) and (\ref{eq:lmata-def}), we
may rewrite equivalently this equation in terms of the total
differential of the ,,improved'' Lagrangian $\lmata$:
\begin{equation}
  \label{eq:warlmata0}
  \begin{split}
    {\rm d} \lmata &= {\rm d} \left( \lmatm - \pi^{\mu\nu} \left(
        {f^\rho}_{\rho\sigma} {f^\sigma}_{\mu\nu} - {f^\rho}_{\mu\sigma}
        {f^\sigma}_{\nu\rho} \right) \right) \\
    &= - \Bigl( \met{K}_{\mu\nu}
    + \met\nabla_\rho {f^\rho_{\mu\nu} - \met\nabla_{(\mu}}
    f^\rho_{\nu)\rho}
    -{f^\rho}_{\sigma\rho}{f^\sigma}_{\mu\nu}
+ {f^\rho}_{\sigma\mu}{f^\sigma}_{\rho\nu}
    \Bigr)  {\rm d} \pi^{\mu\nu} \\
     &\ - \pi^{\mu\nu} {\rm d} \left(
      {f^\rho}_{\sigma\rho}{f^\sigma}_{\mu\nu} -
      {f^\rho}_{\sigma\mu}{f^\sigma}_{\rho\nu} \right)
      + {\met{\dens{J}}_\lambda}^{\mu\nu} {\rm d} \left(
      {\Gamma^\lambda}_{\mu\nu} - {f^\lambda}_{\mu\nu}
      \right) \\
      &\ + p^\kappa {\rm d}
    \varphi_{,\kappa} + {p^\kappa}_{,\kappa} {\rm d} \varphi \ .
  \end{split}
\end{equation}
To calculate $-{\met{\dens{J}}_\lambda}^{\mu\nu} {\rm d}
{f^\lambda}_{\mu\nu}$, we use the equations:
\begin{equation}
  \label{eq:lmata-metric}
  {\met{\dens{J}}_\lambda}^{\mu\nu} =  2 {f^{(\mu}}_{\rho\lambda} \pi^{\nu)\rho} -
    {f^\rho}_{\rho\lambda} \pi^{\mu\nu} -
    \delta^{(\mu}_\lambda {f^{\nu)}}_{\rho\sigma}
      \pi^{\rho\sigma} \ ,
\end{equation}
inverse to~(\ref{eq:f(J,g)}). This implies:
\[
\begin{split}
  - {\met{\dens{J}}_\lambda}^{\mu\nu} {\rm d} {f^\lambda}_{\mu\nu}
  &= - \left( {f^\mu}_{\lambda\rho} \pi^{\rho\nu} +
    {f^\nu}_{\lambda\rho} \pi^{\rho\mu} - {f^\rho}_{\lambda\rho}
    \pi^{\mu\nu} \right) {\rm d}
  {f^\lambda}_{\mu\nu} \\
  &\ \ + \left( {f^\nu}_{\kappa\rho} \pi^{\rho\kappa} +
    {f^\kappa}_{\kappa\rho} \pi^{\nu\rho} - {f^\rho}_{\kappa\rho}
    \pi^{\nu\kappa} \right) {\rm d} {f^\alpha}_{\nu\alpha} \\
  &=
  \pi^{\mu\nu} {\rm d} \left( {f^\rho}_{\sigma\rho}{f^\sigma}_{\mu\nu}
    - {f^\rho}_{\sigma\mu}{f^\sigma}_{\rho\nu} \right) \ .
\end{split}
\]
Inserting this into~(\ref{eq:warlmata0}), we get
\begin{equation}
  \label{eq:warlmata1}
  \begin{split}
    {\rm d} \lmata
    &= - \Bigl(  \met{K}_{\mu\nu}
    + \met\nabla_\rho {f^\rho_{\mu\nu} - \met\nabla_{(\mu}}
    f^\rho_{\nu)\rho}
    -{f^\rho}_{\sigma\rho}{f^\sigma}_{\mu\nu}
+ {f^\rho}_{\sigma\mu}{f^\sigma}_{\rho\nu} \Bigr) {\rm d} \pi^{\mu\nu} \\
     &\ + {\met{\dens{J}}_\lambda}^{\mu\nu} {\rm d}
      {\Gamma^\lambda}_{\mu\nu} + p^\kappa {\rm d}
    \varphi_{,\kappa} + {p^\kappa}_{,\kappa} {\rm d} \varphi \ .
  \end{split}
\end{equation}
Using formula~(\ref{eq:rozklad-Ricci}) we see that the terms in
the bracket combine to $K_{\mu\nu}$. Hence, we obtain the
following, extremely compact formulation of the field equations of the
theory (both the gravitational and matter equations), as a single
equation imposed on the total differential of $\lmata$:
\begin{equation}
  \label{eq:var-lmata}
    {\rm d} \lmata(\pi,\Gamma,\varphi,\partial\varphi)
    = - K_{\mu\nu} {\rm d} \pi^{\mu\nu} +
    {\met{\dens{J}}_\lambda}^{\mu\nu} {\rm d} {\Gamma^\lambda}_{\mu\nu} +
    p^\kappa {\rm d} \varphi_{,\kappa} + {p^\kappa}_{,\kappa} {\rm d}
    \varphi \ .
\end{equation}
This completes the proof of Theorem~\ref{simpl}.

\section{The equivalence of the variational principles}
\label{sec:constr-affine-lagr}

We have shown that one can write the Einstein equations in a much simpler
form $K_{\mu\nu} = -
\frac{\partial\lmata(\pi,\Gamma,\varphi,\partial\varphi)}{\partial\pi^{\mu\nu}}$.
What remains to be shown is that the scalar density $\lmata$ is the affine
matter Lagrangian we've been discussing in
Section~\ref{sec:introduction}. In order to do so, we will show that
it fulfills all dynamical equations generated by the ``true'' affine
matter Lagrangian, taken as a function of $\pi,\Gamma,\varphi$
and $\partial\varphi$ variables.

\begin{theorem}\label{equiv}
  The solutions of the Einstein equations derived from the metric matter
  Lagrangian $\lmatm (\pi, \mconn, \varphi, \partial \varphi)$
  are equivalent to the solutions of the dynamical equations derived from the
  affine matter Lagrangian $\lmata (\pi, \Gamma, \varphi,
  \partial \varphi)$, if and only if
  \[
  {\Gamma^\lambda}_{\mu\nu} = {\mconn^\lambda}_{\mu\nu} + {f^\lambda}_{\mu\nu}
  \]
  and
  \[
  \begin{split}
    \lmata(\pi,\Gamma,\varphi,\partial\varphi)
    &=
    \lmatm(\pi,\mconn(\pi,\Gamma,\varphi,\partial\varphi),\varphi,\partial\varphi) \\
    &\ - \pi^{\mu\nu} \left( {f^\rho}_{\sigma\rho}{f^\sigma}_{\mu\nu} -
      {f^\rho}_{\sigma\mu}{f^\sigma}_{\rho\nu} \right) \ ,
  \end{split}
  \]
  where the \emph{nonmetricity tensor} $f = f(\pi,\Gamma,\varphi,\partial\varphi)$
  is given by the formula
  \[
  \begin{split}
    {f^\lambda}_{\mu\nu} &= \frac{k}{\volfrm} \bigl[
    2{\dens{J}_{(\mu}}^{\rho\lambda}g_{\nu)\rho} -
    {\dens{J}_\rho}^{\rho\lambda}g_{\mu\nu}
     - {\dens{J}_\rho}^{\sigma\tau}g^{\rho\lambda} \left(
      g_{\mu\sigma}g_{\nu\tau} - \tfrac{1}{2}g_{\sigma\tau}g_{\mu\nu}
    \right) \\ &\ + \tfrac{2}{3}
    \delta^\lambda_{(\mu}{\dens{J}_\rho}^{\rho\sigma}g_{\nu)\sigma} -
    \delta^\lambda_{(\mu} {\dens{J}_{\nu)}}^{\rho\sigma}g_{\rho\sigma}
    \bigr] \ ,
  \end{split}
  \]
  with the \emph{nonmetricity current} $\dens{J}$ defined as
  \[
    {\dens{J}_\lambda}^{\mu\nu} := \frac{\partial \lmata
        (\pi,\Gamma,\varphi,\partial\varphi)}{\partial
        {\Gamma^\lambda}_{\mu\nu}} \ .
    \]
Moreover, the above quantity is numerically equal to its metric
analogue: ${\dens{J}_\lambda}^{\mu\nu} =
{\met{\dens{J}}_\lambda}^{\mu\nu}$, where the latter is defined by
the corresponding metric formula~(\ref{eq:metJ-definition}).
\end{theorem}

The proof of the identity ${\dens{J}_\lambda}^{\mu\nu} =
{\met{\dens{J}}_\lambda}^{\mu\nu}$ follows trivially from the
equation~(\ref{eq:var-lmata}). Moreover, the above definition of
the nonmetricity tensor $f$ implies the following identity (see
Appendix~\ref{sec:proof-nonmetricity}), equivalent to equations
(\ref{eq:ELGamma}):
\begin{equation}
  \label{eq:nonmetricity}
  \nabla_\lambda \pi^{\mu\nu} -
  \delta^{(\mu}_\lambda \nabla_\kappa \pi^{\nu)\kappa} =
  {\dens{J}_\lambda}^{\mu\nu} \ .
\end{equation}
Hence, the combined system of the metric field equations for the
``matter + gravity system'', encoded in the
equation~(\ref{eq:var-lmata}), is equivalent to the following set
of equations:
\begin{align}
  \label{eq:dynamical-lmata-Gamma}
  K_{\mu\nu} &= - \frac{\partial \lmata}{\partial \pi^{\mu\nu}} \ , \\
  \label{eq:dynamical-lmata-pi}
  \nabla_\lambda \pi^{\mu\nu} -
  \delta^{(\mu}_\lambda \nabla_\kappa \pi^{\nu)\kappa} &=
  \frac{\partial \lmata}{\partial {\Gamma^\lambda}_{\mu\nu}} \ , \\
  p^\kappa &= \frac{\partial \lmata}{\partial \varphi_{,\kappa}} \ , \\
  {p^\kappa}_{,\kappa} &= \frac{\partial \lmata}{\partial \varphi} \ ,
\end{align}
which are exactly the equations arising from the affine variational
principle (see Appendix~\ref{sec:affine-principle} or
\cite{new-principle}) for the affine matter Lagrangian
$\lmata$.

The argument given above can be reversed, constructing the metric
matter Lagrangian from a given affine matter Lagrangian. The definition
(\ref{eq:f(J,g)}) arises then as the solution of the Euler-Lagrange
equations~(\ref{eq:dynamical-lmata-pi}) for $\Gamma$. This completes
the proof of Theorem~\ref{equiv}.

Adding to $\lmata$ the term $\pi^{\mu\nu} K_{\mu\nu}$, where we
insert the relation $\pi=\pi(K,\Gamma,\phi,\partial\phi)$ given
implicitly by equations~(\ref{eq:dynamical-lmata-Gamma}), we define
the total affine Lagrangian:
\begin{equation}
  \label{eq:laff-from-lmata}
    \laff (K, \Gamma, \varphi, \partial\varphi ) := \lmata(\pi,\Gamma,\varphi,\partial\varphi) +
    \pi^{\mu\nu}(K,\Gamma,\varphi,\partial\varphi) K_{\mu\nu} \ .
  \end{equation}
This is a Legendre transformation which converts
equation~(\ref{eq:var-lmata}) to the following, equivalent
equation
\begin{equation}
  \label{eq:var-laff-pocz}
  {\rm d} \laff(K,\Gamma,\varphi,\partial\varphi)
  = \pi^{\mu\nu} {\rm d} K_{\mu\nu}   +
  {\dens{J}_\lambda}^{\mu\nu} {\rm d} {\Gamma^\lambda}_{\mu\nu} +
p^\kappa {\rm d} \varphi_{,\kappa} + {p^\kappa}_{,\kappa} {\rm d}
\varphi \ .
\end{equation}
These are precisely (see Appendix \ref{sec:affine-principle}) the
Euler-Lagrange equations derived in the affine formulation of the
theory.

This completes the proof of the equivalence of the metric and
the affine principles: each solution of the dynamical equations
generated by the metric principle can be transformed into the
corresponding solution of the dynamical equations generated by the
affine principle.

It is interesting to observe that the numerical value of the total
affine Lagrangian is equal to
\begin{equation}
\begin{split}
\laff &= \lmatm + \pi^{\mu\nu} \left( K_{\mu\nu} -
    {f^\rho}_{\sigma\rho}{f^\sigma}_{\mu\nu} +
    {f^\rho}_{\sigma\mu}{f^\sigma}_{\rho\nu} \right) \\
&= \lmet + \pi^{\mu\nu} \bigl( K_{\mu\nu} - \met{K}_{\mu\nu} -
    {f^\rho}_{\sigma\rho}{f^\sigma}_{\mu\nu} +
    {f^\rho}_{\sigma\mu}{f^\sigma}_{\rho\nu} \bigr) \\
&= \lmet + \pi^{\mu\nu} \left( \met\nabla_\lambda
      {{f^\lambda}_{\mu\nu}
        - \met\nabla_{(\mu} f^\rho}_{\nu)\rho} \right) \\
&= \lmet + \partial_\lambda \left( \pi^{\mu\nu} \left(
        {f^\lambda}_{\mu\nu} - \delta^\lambda_{\mu}
        {f^\rho}_{\nu\rho} \right) \right) \ ,
\end{split}
\end{equation}
and, whence, differs from the total metric Lagrangian by a
complete divergence only (see also \cite{Torino}).

The transformation from the metric to the affine picture may be
summarized as follows: we rewrite the numerical value of $\laff$
as:
\begin{equation}
 \label{eq:num-laff} \laff = \lmatm - \pi^{\mu\nu} \frac{\partial
 \lmatm(\pi,\mconn,\varphi,\partial\varphi)}{\partial \pi^{\mu\nu}}
 \ .
\end{equation}
Inserting~\eqref{eq:f(J,g)} into the equation $\Gamma = \mconn +
f(\pi,\met\Gamma,\varphi,\partial\varphi)$, and solving the
resulting equation with respect to $\mconn$ (where
$\met{\dens{J}}$ is treated as a function of $\pi,\mconn,\varphi$
and $\partial\varphi$), we get the relation $\mconn =
\mconn(\pi,\Gamma,\varphi,\partial\varphi)$. Inserting this
relation into $f = f(\pi,\mconn,\varphi,\partial\varphi)$ we get
$f = f(\pi,\Gamma,\varphi,\partial\varphi)$. Inserting this
relation into the modified version~\eqref{eq:Einst5} of the metric
Einstein equations:
\begin{equation}
 \label{eq:Einst-6} K_{\mu\nu} - {f^\rho}_{\mu\nu}
 {f^\sigma}_{\rho\sigma} + {f^\rho}_{\mu\sigma}
 {f^\sigma}_{\nu\rho} = -
 \frac{\partial\lmatm(\pi,\mconn(\pi,\Gamma,\varphi,\partial\varphi)
 ,\varphi,\partial\varphi)}{\partial
 \pi^{\mu\nu}}
 \end{equation}
we get the relation $K = K(\pi,\Gamma,\varphi,\partial\varphi)$.
Inverting this relation with respect to the metric we get $\pi =
\pi(K,\Gamma,\varphi,\partial\varphi)$. Inserting this relation
into the relation $\mconn =
\mconn(\pi,\Gamma,\varphi,\partial\varphi)$ gives the relation
$\mconn = \mconn(K,\Gamma,\varphi,\partial\varphi)$. We use these
two relations to eliminate the metric quantities $\pi$ and $\mconn$
from the right-hand side of~\eqref{eq:num-laff}. This way we
obtain $\laff= \laff(K , \Gamma , \varphi , \partial\varphi)$
without computing the intermediate quantity $\lmata$.

For the matter Lagrangians obeying the condition~(\ref{eq:part-fi})
(i.~e.~for the theories considered in~\cite{new-principle}), we have
$f\equiv 0$ and most of the above relations are trivial:
$\Gamma=\mconn$ and $K = \met{K}$. The only relation which remains
to be inverted is the Einstein equation $K =
K(\pi,\varphi,\partial\varphi)$. The resulting relation $\pi =
\pi(K,\varphi,\partial\varphi)$, when inserted into the metric
Lagrangian $\lmet$, gives us $\laff$. \normalfont

\section{Conclusions}
\label{sec:conclusions}

The affine variational principle leads to such a formulation of the General
Relativity theory which is, in some aspects, simpler than the one
based on the metric principle. The metric formulation has the
following problems:
\begin{enumerate}
\item The Hilbert Lagrangian is of second differential
order in $g$, and
  yet it generates second order (and not 4-th order, as one would expect)
  equations. This is due to nongeneric cancellations.
\item The status of the momentum canonically conjugate to the
metric is obscure: if one takes as the momentum the quantity
  $\frac{\partial \lhilb}{\partial g_{\mu\nu,\lambda\kappa}}$, which is
  formally correct, one gets
  second type constraints for the canonical momentum:
  \[
  \frac{\partial \lhilb}{\partial g_{\mu\nu,\lambda\kappa}} \equiv
  \frac{\sqrt{|\det g}}{2k} \left( g^{\mu\kappa} g^{\nu\lambda} -
    g^{\mu\nu} g^{\lambda\kappa} \right)\ .
  \]
The correct canonical structure (the ADM structure) is obtained
only when we perform the symplectic reduction of the theory with
respect to the above constraints.
\item As already mentioned, the
``sources of gravity'',
    represented in the metric formulation by the energy-momentum tensor,
    contain not only the first but also the second derivatives of the metric
    and, therefore, may change
    completely the causal behaviour of the Einstein equations.
    The second order term contained in the Einstein tensor cannot be considered
    anymore as the ``principal part'' of the equations because of another
    second order term contained in the metric energy-momentum tensor.
\end{enumerate}
The affine principle is free of these defects. The affine Lagrangian
$\laff$ is of the first differential order in $\Gamma$ and,
correctly, generates second order Euler-Lagrange equations.  No
additional constraints on the symmetric connection $\Gamma$ are
imposed. No total derivatives are ``built in'' in the Lagrangian,
and it is not required to be of any special form. The canonical
momentum, given by formula~(\ref{eq:pi4}), is well defined,
transforms like a tensor and represents another geometric quantity
describing the geometric structure of space-time: the metric tensor.
The canonical structure of the theory, first described by
Arnowitt, Deser and Misner in their fundamental paper~\cite{ADM},
considerably simplifies in this formulation (see
\cite{canonical-structure} for details). Finally, as shown by
Ashtekar-Lewandowski approach to quantum gravity (see
\cite{ashtekar,ashtekar96,qgrav-srep}),
the description of the configuration of the gravitational field in
terms of the connection is much more natural. Contrary to the
space of metric tensors, the space of connections carries an
affine structure and provides a natural arena for path integrals.

\section*{Appendixes}

\appendix

\section{The affine variational principle}
\label{sec:affine-principle}

In \cite{new-principle} a class of variational principles has been
considered, based on the affine, symmetric (not necessarily metric)
connection $\Gamma$ as the configuration variable. The Lagrangian was
assumed to be an invariant scalar density built from $\Gamma$ and its
first partial derivatives $\partial \Gamma$, together with the matter
field(s) $\varphi$ and its first partial derivatives:
\begin{equation}
\label{eq:laff-grav} \laff = \laff(\Gamma, \partial\Gamma,
\varphi, \partial \varphi)\ .
\end{equation}
Defining the canonical momenta: conjugate to $\Gamma$
\begin{equation}\label{eq:pi4}
    {\pi_\lambda}^{\mu\nu\kappa} := \frac{\partial
    \laff(\Gamma,\partial\Gamma,\varphi,\partial\varphi)}{\partial
    {\Gamma^\lambda}_{\mu\nu,\kappa}} \ ,
\end{equation}
where ${\Gamma^\lambda}_{\mu\nu,\kappa} =\partial_\kappa
{\Gamma^\lambda}_{\mu\nu}$, and conjugate to $\varphi$
\begin{equation}
  \label{eq:matt-can-mom}
  p^\kappa := \frac{\partial \laff (\Gamma, \partial\Gamma,
      \varphi, \partial\varphi)}{\partial \varphi_{,\kappa}} \ ,
\end{equation}
we calculate the variation of the Lagrangian as follows:
\begin{equation}
  \label{eq:delta-laff}
  \begin{split}
  {\rm d} \laff   &= {\pi_\lambda}^{\mu\nu\kappa} {\rm d}
  {\Gamma^\lambda}_{\mu\nu,\kappa}   +
  \frac{\partial \laff}{\partial
    {\Gamma^\lambda}_{\mu\nu}}
     {\rm d} {\Gamma^\lambda}_{\mu\nu} +
   p^\kappa {\rm d} \varphi_{,\kappa} +
   \frac{\partial \laff}{\partial \varphi} {\rm d} \varphi
   \\
   &= \left(
  \frac{\partial \laff}{\partial
    {\Gamma^\lambda}_{\mu\nu}} -
    \partial_\kappa {\pi_\lambda}^{\mu\nu\kappa} \right)
     {\rm d} {\Gamma^\lambda}_{\mu\nu} +
   \left(
   \frac{\partial \laff}{\partial \varphi} -
   \partial_\kappa p^\kappa \right) {\rm d} \varphi
   \\ &\ + \partial_\kappa \left( {\pi_\lambda}^{\mu\nu\kappa}
   {\rm d} {\Gamma^\lambda}_{\mu\nu} + p^\kappa {\rm d} \varphi
   \right)
   \ .
 \end{split}
\end{equation}
The Euler-Lagrange equations of the theory:
\begin{equation}
  \label{eq:EL-aff-1-short}
  \begin{split}
    \partial_\kappa {\pi_\lambda}^{\mu\nu\kappa} &= \frac{\partial
      \laff \left( \Gamma, \partial\Gamma, \varphi, \partial\varphi
      \right)}{\partial {\Gamma^\lambda}_{\mu\nu}} \\
    \partial_\kappa p^\kappa &= \frac{\partial \laff (\Gamma,
      \partial\Gamma, \varphi, \partial\varphi)}{\partial \varphi}
  \end{split}
\end{equation}
mean that the volume part of the variation must vanish or,
equivalently, the entire variation ${\rm d} \laff$ reduces to the
boundary part:
\begin{align}
\label{bound} \mathrm{d}\laff &= \partial_\kappa \left( {\pi_\lambda}^{\mu\nu\kappa}
   {\rm d} {\Gamma^\lambda}_{\mu\nu} + p^\kappa {\rm d} \varphi
   \right) \\
\label{boundary} &= \begin{aligned}[t]
&{\pi_\lambda}^{\mu\nu\kappa} {\rm d}
  {\Gamma^\lambda}_{\mu\nu,\kappa}
  + (\partial_\kappa {\pi_\lambda}^{\mu\nu\kappa} )
   {\rm d} {\Gamma^\lambda}_{\mu\nu} \\
&+
   p^\kappa {\rm d} \varphi_{,\kappa} +
   (\partial_\kappa p^\kappa ) {\rm d} \varphi
   \ .
                    \end{aligned}
\end{align}

To obtain a theory which is invariant with respect to
diffeomorphisms we assume that $\laff$ depends upon the derivatives of
$\Gamma$ {\em via} the only invariant quantity which one may
construct: the Riemann curvature tensor $R^\lambda_{\
\mu\nu\kappa}$. Lagrangians of the type
$\laff=\laff(\Gamma,R,\varphi,\partial\varphi)$ describe a much
broader class of theories than the General Relativity theory (see
\cite{ferraris1,ferraris2}), i.~e.~the resulting field
equations are not necessarily equivalent to the Einstein equations.
However, it has been proved in~\cite{new-principle} that we can
reduce this freedom to exactly Einstein's theory by postulating a
specific dependence of $\laff$ upon the Riemann tensor: the only
information about the curvature which the Lagrangian is sensitive
to is the symmetric part $K_{\mu\nu}= \tfrac 12
(R_{\mu\nu}+R_{\nu\mu})$ of its Ricci tensor
$R_{\mu\nu}=R^\lambda_{\ \mu\lambda\nu}$ (unlike in the case of the
metric connection, the Ricci tensor is not necessarily symmetric):
\begin{equation}
\label{eq:laff-grav-K} \laff = \laff(\Gamma, K, \varphi, \partial
\varphi).
\end{equation}
Hence, the variation~\eqref{eq:delta-laff} of the Lagrangian
reduces to
\begin{equation}\label{eq:delta-laff-K}
    {\rm d} \laff   = \pi^{\mu\nu} {\rm d}
  K_{\mu\nu}   +
  {\dens{J}_\lambda}^{\mu\nu}
     {\rm d} {\Gamma^\lambda}_{\mu\nu} +
   p^\kappa {\rm d} \varphi_{,\kappa} +
   \frac{\partial \laff}{\partial \varphi} {\rm d} \varphi \ ,
\end{equation}
where we have introduced the following tensor densities:
\begin{equation}
  \label{eq:defpi-bzz}
  \pi^{\mu\nu} := \frac{\partial \laff(\Gamma, K, \varphi,
  \partial \varphi)}{\partial K_{\mu\nu}} \ ,
\end{equation}
and
\begin{equation}
  \label{eq:J-def-bzz}
  {\dens{J}_\lambda}^{\mu\nu} := \frac{\partial \laff
      (\Gamma, K, \varphi, \partial\varphi)}{\partial
      {\Gamma^\lambda}_{\mu\nu}} \ .
\end{equation}

However, $K$ is linear in $\partial\Gamma$ and quadratic in $\Gamma$:
\begin{equation}\label{Ka}
        K_{\mu\nu} = {\Gamma^\lambda}_{\mu\nu,\lambda} -
    {\Gamma^\lambda}_{\lambda(\mu, \nu)} +
    {\Gamma^\lambda}_{\alpha\lambda} {\Gamma^\alpha}_{\mu\nu}
    -{\Gamma^\lambda}_{\alpha\nu} {\Gamma^\alpha}_{\mu\lambda}
    \ .
\end{equation}
Hence, the first term in~\eqref{eq:delta-laff-K} may be calculated
as follows:
\begin{equation}
\label{pi-delta-K}
\begin{split}
  \pi^{\mu\nu} {\rm d} K_{\mu\nu} &=
  \left( \delta^\kappa_\lambda \pi^{\mu\nu} -
    \delta_\lambda^{(\mu} \pi^{\nu)\kappa} \right)
  {\rm d} {\Gamma^\lambda}_{\mu\nu,\kappa}  \\
  &\ +\left(\pi^{\mu\nu} {\Gamma^\kappa}_{\lambda\kappa} +
    \pi^{\sigma\kappa} {\Gamma^\mu}_{\sigma\kappa}\delta^\nu_\lambda
    - 2 \pi^{\sigma\nu} {\Gamma^\mu}_{\sigma\lambda}
  \right) {\rm d} {\Gamma^\lambda}_{\mu\nu} \ .
\end{split}
\end{equation}
Comparing it with~\eqref{boundary} we obtain from~(\ref{eq:delta-laff-K}):
\begin{equation}  \label{eq:pipi2}
    {\pi_\lambda}^{\mu\nu\kappa}  =
    \delta^\kappa_\lambda \pi^{\mu\nu} - \delta_\lambda^{(\mu}
    \pi^{\nu)\kappa} \ .
\end{equation}
Taking into account that $\pi$ is a {\em tensor density}, it is
easy to check that~\eqref{pi-delta-K} contains the difference
between the partial and the covariant divergence:
\begin{equation}\label{partial-nabla}
    \pi^{\mu\nu} {\Gamma^\kappa}_{\lambda\kappa} +
    \pi^{\sigma\kappa} {\Gamma^\mu}_{\sigma\kappa}\delta^\nu_\lambda
    - 2 \pi^{\sigma\nu} {\Gamma^\mu}_{\sigma\lambda} =
    \partial_\kappa {\pi_\lambda}^{\mu\nu\kappa} -
    \nabla_\kappa {\pi_\lambda}^{\mu\nu\kappa} \ .
\end{equation}
Combining these results, the variation~\eqref{eq:delta-laff-K}
reduces to
\begin{equation}
  \begin{split}
    \label{eq:delta-laff-K1}
    {\rm d} \laff  &=
    \left({\dens{J}_\lambda}^{\mu\nu} -
      \nabla_\kappa {\pi_\lambda}^{\mu\nu\kappa} \right)
    {\rm d} {\Gamma^\lambda}_{\mu\nu} +
    \left(
      \frac{\partial \laff}{\partial \varphi} -
      \partial_\kappa p^\kappa \right) {\rm d} \varphi \\
    &\ + \partial_\kappa \left( {\pi_\lambda}^{\mu\nu\kappa}
      {\rm d} {\Gamma^\lambda}_{\mu\nu} + p^\kappa {\rm d} \varphi
    \right)\ .
  \end{split}
\end{equation}
The Euler-Lagrange equations mean that the volume part
of~\eqref{eq:delta-laff-K1} vanishes:
\begin{equation}
  \label{eq:EL-aff-2-cov-a}
  \nabla_\kappa {\pi_\lambda}^{\mu\nu\kappa} =
  \nabla_\lambda \pi^{\mu\nu} -
  \delta_\lambda^{(\mu} \nabla_\rho \pi^{\nu)\rho} =
  {\dens{J}_\lambda}^{\mu\nu}\ ,\quad
  {p^\kappa}_{,\kappa} = \frac{\partial \laff }{\partial \varphi} \ .
\end{equation}
We interpret $\pi$ as the contravariant density of the metric:
  \begin{equation}
    \label{eq:pimetric}
    \pi^{\mu\nu} =: \frac{\sqrt{|\det g|} }{2k} \ g^{\mu\nu} \ .
  \end{equation}
The formula ${\pi_\rho}^{\mu\nu\rho} = 4 \pi^{\mu\nu} - \pi^{\mu\nu}$
or, equivalently,
\begin{equation}
\label{eq:pipi} \pi^{\mu\nu}  = \frac{1}{3}
{\pi_\rho}^{\mu\nu\rho} \ ,
\end{equation}
gives a one-to-one correspondence between
${\pi_\lambda}^{\mu\nu\kappa}$ and $\pi^{\mu\nu}$. Taking the trace of the
first equation of~\eqref{eq:EL-aff-2-cov-a}, we obtain the relation
${\dens{J}_\rho}^{\mu\rho} = \nabla_\rho \pi^{\mu\rho} - \frac{5}{2}
\nabla_\rho \pi^{\mu\rho}$, which leads to an equivalent form of
the Euler-Lagrange equations generated by $\laff$:
\begin{equation}
  \label{eq:EL-aff-2-cov-b}
  \nabla_\lambda \pi^{\mu\nu} = {\dens{J}_\lambda}^{\mu\nu} -
  \tfrac{2}{3} \delta^{(\mu}_\lambda {\dens{J}_\rho}^{\nu)\rho}
     \quad
    {p^\kappa}_{,\kappa} = \frac{\partial \laff (\Gamma, \partial\Gamma, \varphi,
    \partial\varphi)}{\partial \varphi}
\end{equation}
If $\laff= \laff (K, \varphi, \partial\varphi)$ (which was assumed
in the paper \cite{new-principle}) then, according to~\eqref{eq:J-def-bzz},
$\dens{J} = 0$ and the gravitational part of
these equations becomes
\[
\nabla_\lambda \pi^{\mu\nu} = 0 \ .
\]
The only solution $\Gamma$ of this equations is the metric
(Levi-Civitta) connection: $\Gamma = \mconn$. In a generic case,
the only solution of the ``nonmetricity equations''
$\nabla_\lambda \pi^{\mu\nu} = {\dens{J}_\lambda}^{\mu\nu} -
\tfrac{2}{3} \delta_\lambda^{(\mu} {\dens{J}_\rho}^{\nu)\rho}$ is
the following connection:
\[
\Gamma:= \mconn + f \ ,
\]
where the
``nonmetricity tensor'' $f$ is uniquely defined in terms of the
``nonmetricity current'' $\dens{J}$ by formula~(\ref{eq:f(J,g)})
(see Appendix \ref{sec:proof-nonmetricity} for the proof).

To analyze the relation of the above ``affine theory'' with the
conventional metric Einstein theory it is useful to subtract from
the Lagrangian $\laff$ the generalized Hilbert term ${\cal L}_{\rm H}$,
given by the scalar curvature:
\begin{equation}\label{Hilb-gen}
    {\cal L}_{\rm H}=\volfrmfrac R =
    \volfrmfrac g^{\mu\nu} R_{\mu\nu} = \pi^{\mu\nu}
     K_{\mu\nu} \ .
\end{equation}
This way we define the ``affine matter Lagrangian'':
\begin{equation}
  \label{eq:Legendre-laff-to-lmata}
  \begin{split}
    \lmata (\pi, \Gamma, \varphi, \partial\varphi) &:= \laff(\Gamma,
    K(\pi,\Gamma,\varphi,\partial\varphi), \varphi, \partial \varphi)
    \\
    &\ - \pi^{\mu\nu} K_{\mu\nu} \left(\pi,\Gamma,\varphi,\partial\varphi
    \right) \ .
\end{split}
\end{equation}
The above transformation may be interpreted as the Legendre
transformation between $\pi^{\mu\nu}$ and $K_{\mu\nu}$. To rewrite
the field equations of the theory in terms of $\lmata$, we use the standard
technique of the Legendre transformation~\cite{framework,ham-field-theory}.
For this purpose we write the
field equations~\eqref{eq:EL-aff-2-cov-a} together with the
definitions~\eqref{eq:matt-can-mom} and~\eqref{eq:defpi-bzz} of
the momenta in the form of a generating relation imposed on the
total derivative of $\laff$:
\begin{equation}
  \label{Gen-Lag}
  \begin{split}
    {\rm d} \laff(K,\Gamma,\varphi,\partial\varphi)
    &= \pi^{\mu\nu} {\rm d} K_{\mu\nu} +
    p^\kappa {\rm d} \varphi_{,\kappa} + {p^\kappa}_{,\kappa} {\rm d}
    \varphi \\ &\ + \Bigl(
      \nabla_\lambda \pi^{\mu\nu} -
      \delta_\lambda^{(\mu} \nabla_\rho \pi^{\nu)\rho} \Bigr)
    {\rm d} {\Gamma^\lambda}_{\mu\nu} \ .
  \end{split}
\end{equation}
This implies the following equation for the total differential of the
function \eqref{eq:Legendre-laff-to-lmata}:
\begin{equation}
  \label{Gen-Ham}
  \begin{split}
    &{\rm d} \lmata (\pi, \Gamma, \varphi, \partial\varphi)
    = - K_{\mu\nu}  {\rm d}  \pi^{\mu\nu}
    + \left( \nabla_\lambda \pi^{\mu\nu} - \delta_\lambda^{(\mu}
      \nabla_\rho \pi^{\nu)\rho} \right) {\rm d}
    {\Gamma^\lambda}_{\mu\nu} \\
&\ + p^\kappa {\rm d} \varphi_{,\kappa} +
    {p^\kappa}_{,\kappa} {\rm d} \varphi \ ,
 \end{split}
\end{equation}
or, equivalently:
\begin{equation}
  \label{eq:EL-lmata-1}
  \begin{split}
    K_{\mu\nu} &= - \frac{\partial
      \lmata(\pi,\Gamma,\varphi,\partial\varphi)}{\partial
      \pi^{\mu\nu}} \\
 {\dens{J}_\lambda}^{\mu\nu} &= \nabla_\lambda \pi^{\mu\nu} -
    \delta_\lambda^{(\mu} \nabla_\rho \pi^{\nu)\rho} \\
    {p^\kappa}_{,\kappa} &= \frac{\partial \lmata ( \pi, \Gamma, \varphi,
    \partial\varphi )}{\partial \varphi} \ ,
  \end{split}
\end{equation}
where the quantities $\dens{J}$ and $p$ are now defined as
\begin{equation}
  \label{eq:densJ-lmata}
  {\dens{J}_\lambda}^{\mu\nu} := \frac{\partial \lmata
    (\pi,\Gamma,\varphi,\partial\varphi)}{\partial {\Gamma^\lambda}_{\mu\nu}}
\end{equation}
and
\begin{equation}
  \label{eq:mattmom-lmata}
  p^\kappa := \frac{\partial \lmata
    (\pi,\Gamma,\varphi,\partial\varphi)}{\partial \varphi_{,\kappa}}
  \ .
\end{equation}
Of course, the equivalence between equations~\eqref{eq:EL-lmata-1} and
\eqref{eq:EL-aff-2-cov-b} may be proved directly\footnote{The
  proof based on formula~\eqref{Gen-Ham} uses, in fact, symplectic relations.
  It is valid also when there are Lagrangian constraints imposed
  on the configuration variables. In such a theory the ``response'' of the
  system is not unique: the relation $K=K(\pi,\Gamma,\varphi,\partial\varphi)$
  may be noninvertible, but still formula~\eqref{eq:Legendre-laff-to-lmata}
  implies the field equations~\eqref{eq:EL-lmata-1} (see~\cite{framework,ham-field-theory} for the
  details).} by calculating the derivatives of $\lmata$ defined by~\eqref{eq:Legendre-laff-to-lmata}.

We will now show that the first equations of~\eqref{eq:EL-lmata-1}
may be rewritten in the conventional form:
\begin{equation}\label{eq:T-aff}
    \dens{G}^{\mu\nu} =  k \dens{T}^{\mu\nu} \ ,
\end{equation}
where the ,,affine energy-momentum tensor density'' $\dens{T}$ is
defined as follows:
\begin{equation}
  \label{eq:Einst-eq-lmata-pre}
  \dens{T}^{\mu\nu} = 2  \frac{\partial \lmata(g,\Gamma, \varphi,
    \partial\varphi)}{\partial g_{\mu\nu}} \ ,
\end{equation}
and
\[
\dens{G}^{\mu\nu}:=\sqrt{\abs{\det g}}(K^{\mu\nu} - \tfrac 12
g^{\mu\nu} R)
\]
is the density of the Einstein tensor constructed
from $K$. Indeed, according to \eqref{eq:pimetric}, the metric tensor
$g$ can be reconstructed from $\pi$ as
\begin{equation}
  \label{eq:g-od-pi}
  g_{\mu\nu} = 2k \sqrt{\abs{\det \pi}} ( \pi^{-1} )_{\mu\nu} \ .
\end{equation}
Inserting~\eqref{eq:g-od-pi} into the first equations of~\eqref{eq:EL-lmata-1},
we get
\[
\begin{split}
  K_{\mu\nu} &= - \frac{\partial g_{\rho\sigma}}{\partial
    \pi^{\mu\nu}} \frac{\partial \lmata (g, \Gamma, \varphi,
    \partial\varphi)}{\partial
    g_{\rho\sigma}} \\
  &= 2k \sqrt{\abs{\det \pi}} \biggl( ( \pi^{-1} )_{\rho(\mu}
    (\pi^{-1})_{\nu)\sigma}
- \frac{1}{2} (\pi^{-1})_{\mu\nu}
    (\pi^{-1})_{\rho\sigma} \biggr) \\
&\ \times \frac{\partial \lmata (g, \Gamma,
    \varphi, \partial\varphi)}{\partial g_{\rho\sigma}} \\
  &= \frac{2k}{\sqrt{\abs{\det g}}} \left( g_{\rho(\mu} g_{\nu)\sigma}
    - \frac{1}{2} g_{\mu\nu} g_{\rho\sigma} \right)
\frac{\partial
    \lmata (g, \Gamma, \varphi, \partial\varphi)}{\partial
    g_{\rho\sigma}} \ ,
\end{split}
\]
Inverting this equations we obtain~\eqref{eq:T-aff}.

In the paper~\cite{Mauro} generalized conservation laws for the
energy-momentum tensor~\eqref{eq:Einst-eq-lmata-pre} (matching the
generalized Bianchi identities for the nonmetric Einstein tensor
$\dens{G}$) were analyzed. In the special case of $\dens{J} = 0$ we
have $\Gamma=\mconn$ and ${\cal T}$ reduces to the standard
(metric) energy-momentum tensor, provided we identify $\lmata$
with the (metric) matter Lagrangian $\lmatm$. Moreover,~(\ref{eq:Einst-eq-lmata-pre})
become the conventional Einstein
equations for a matter Lagrangian independent of $\partial g$:
\begin{equation}
  \label{eq:Einst-eq-lmata}
  \met{\dens{G}}^{\mu\nu} = 2 k \frac{\partial \lmatm(g, \varphi,
    \partial\varphi)}{\partial g_{\mu\nu}} \ .
\end{equation}
In this way the equivalence between the affine and the metric approaches
was proved in~\cite{new-principle} in the special case $\dens{J} =
0$. The only difference between the two principles in this case is
that in the affine approach, equations~\eqref{eq:Einst-eq-lmata} are
{\em postulated} (by the specific choice of the Lagrangian) and the
metricity equation is a consequence of the
Euler-Lagrange equations~\eqref{eq:EL-aff-2-cov-b}, while in
the metric approach the metricity is {\em postulated}
and~\eqref{eq:Einst-eq-lmata} (or~\eqref{eq:defpi-bzz})
is a consequence of the Euler-Lagrange equations.

For the purposes of canonical relativity, let us observe that the field
dynamics, written in terms of equation~\eqref{bound}, may be
further reduced. Indeed, the identities~\eqref{eq:pipi2} imply:
\begin{equation} \label{pidR}
  \partial_\kappa \left( {\pi}_{\lambda}^{\ \mu\nu\kappa} {\rm d}
  {\Gamma}^{\lambda}_{\mu\nu} \right) =
    \partial_\lambda \left( {\pi}^{\mu\nu} {\rm d}
  A^{\lambda}_{\mu\nu} \right)   \ ,
\end{equation}
where we have denoted
\begin{equation}
  A^{\lambda}_{\mu\nu} := {\Gamma}^{\lambda}_{\mu\nu} -
  {\delta}^{\lambda}_{(\mu} {\Gamma}^{\kappa}_{\nu ) \kappa} \ .
\end{equation}
Hence, equation~\eqref{bound} may be written as follows:
\begin{equation}
    {\rm d} \laff = \partial_\lambda \left( {\pi}^{\mu\nu} {\rm d}
  A^{\lambda}_{\mu\nu}  + p^\lambda {\rm d} \varphi
   \right) \ .
\end{equation}
Integrating it over a four-dimensional volume ${\cal O}$ we obtain
\begin{equation}\label{c}
{\rm d} \int_{\cal O} \laff =  \int_{\partial{\cal O}}
  {\pi}^{\mu\nu} {\rm d}
  A^{\perp}_{\mu\nu} + p^\perp {\rm d} \varphi \ .
\end{equation}
It turns out that not all among the 10 components
$A^{\perp}_{\mu\nu}$ transversal to the three-surface
${\partial{\cal O}}$ are independent. It was proved in
\cite{canonical-structure} that the four components
$A^{\perp}_{\mu\perp}$ may be calculated in terms of the 6
components $A^{\perp}_{kl}$, where $k$ and $l$ label the three
coordinates on the boundary. Reducing the right side of~\eqref{c}
with respect to these constraints, we obtain the
following form of dynamics:
\begin{equation}
{\rm d} \int_{\cal O} \laff  = - \frac 1{16 \pi}
\int_{\partial{\cal O}} g_{kl} \;  \delta \Pi^{kl}
  +\int_{\partial{\cal O}}
   p^\perp {\rm d} \varphi \ ,
\label{dL-overO}
\end{equation}
where $g_{kl}$ is the three-metric of ${\cal O}$ and $\Pi^{kl}$
denotes its external curvature in the ADM representation (see
\cite{canonical-structure}). Suppose that ${\cal O} = V \times
[t_1 , t_2 ]$, where $V$ is three-dimensional and spatial, and that $[t_1 ,
t_2 ]$ is the time interval. Then, the boundary ${\partial{\cal
O}}$ splits into the space-like part composed of two surfaces
$\Sigma_i = V \times \{t_i\}$ and the time-like world-tube
$T=\partial V \times [t_1 , t_2 ]$. The integration
of~\eqref{dL-overO} over the Cauchy surfaces $\Sigma_i$ gives just the
well known ADM canonical structure. But the remaining integral
over $T$ captures all the boundary terms which are necessary for
the definition of the gravitational Hamiltonian. Moreover, there
is also a Dirac-delta-like contribution of the external curvature
corresponding to the corners $\partial V \times \{t_i\}$. This
contribution modifies the ADM canonical structure within the
closed volume $V$ and makes it gauge invariant (see again
\cite{canonical-structure})

\section{Proofs of various formulas}
\label{sec:proofs-vari-form}

\subsection{The proof of formula (\ref{eq:Einst3})}
\label{proof}

The matter Lagrangian depends upon the metric \textit{via} the
metric itself and the metric connection coefficients:
\begin{equation}
  \label{eq:term1}
  \begin{split}
  &\frac{\partial \lmatm(g,\partial g, \varphi, \partial\varphi)}
  {\partial g_{\mu\nu}} \\
  &=
  \frac{\partial \lmatm(g,\mconn, \varphi, \partial\varphi)}
  {\partial g_{\mu\nu}} +
  \frac{\partial {\mconn^\alpha}_{\beta\gamma}}{\partial g_{\mu\nu}}
  \frac{\lmatm (g, \mconn, \varphi, \partial\varphi)}
  {\partial {\mconn^\alpha}_{\beta\gamma}} \\
  &= \frac{\partial \lmatm(g,\mconn, \varphi, \partial\varphi)}
  {\partial g_{\mu\nu}}
  - \frac{1}{2} g^{\alpha(\mu} g^{\nu)\delta} \left(
    g_{\delta\beta,\gamma} + g_{\delta\gamma,\beta} -
    g_{\beta\gamma,\delta} \right) {\met{\dens{J}}_\alpha}^{\beta\gamma} \\
  &= \frac{\partial \lmatm(g,\mconn, \varphi, \partial\varphi)}
  {\partial g_{\mu\nu}} -
  g^{\alpha(\mu} {\mconn^{\nu)}}_{\beta\gamma}
  {\met{\dens{J}}_\alpha}^{\beta\gamma} \ ,
\end{split}
\end{equation}
where the tensor density $\met{\dens{J}}$ defined by
\begin{equation}
  \label{eq:def-J}
  {\met{\dens{J}}_\lambda}^{\mu\nu} := \frac{\partial \lmatm
    \left(g,\mconn,\varphi,\partial\varphi\right)}{\partial
    {\mconn^\lambda}_{\mu\nu}}
\end{equation}
is numerically equal to the nonmetricity current $\dens{J}$ in the
affine formulation, see~formula~(\ref{eq:var-lmata}).

The second term on the right side of~(\ref{eq:Einst3}) may also
be rewritten along these lines:
\begin{equation}
 \label{eq:term2}
\begin{split}
  \partial_\lambda \frac{\partial \lmatm(g,\partial
    g,\varphi,\partial\varphi)}{\partial g_{\mu\nu,\lambda}} &=
  \partial_\lambda \frac{\partial {\mconn^\alpha}_{\mu\nu}}{\partial
    g_{\mu\nu,\lambda}} \frac{\partial
    \lmatm(g,\mconn,\varphi,\partial\varphi)}{\partial
    {\mconn^\alpha}_{\beta\gamma}} \\
  &= \frac{1}{2} \Bigl[ 2 g^{\alpha(\mu} \partial_\gamma
  {\met{\dens{J}}_\alpha}^{\nu)\gamma} - g^{\alpha\gamma}
  \partial_\gamma {\met{\dens{J}}_\alpha}^{\mu\nu} \\
  &\ - \left( 2 g^{\alpha\rho} g^{\sigma(\mu}
    {\met{\dens{J}}_\alpha}^{\nu)\gamma} - g^{\alpha\rho}
    g^{\gamma\sigma} {\met{\dens{J}}_\alpha}^{\mu\nu} \right)
  g_{\rho\sigma,\gamma} \Bigr] \\
  &= \frac{1}{2} \Bigl[ 2 g^{\alpha(\mu} \partial_\beta
  {\met{\dens{J}}_\alpha}^{\nu)\beta} - g^{\alpha\beta} \partial_\beta
  {\met{\dens{J}}_\alpha}^{\mu\nu} - 2 g^{\alpha\beta}
  {\mconn^{(\mu}}_{\beta\gamma}
  {\met{\dens{J}}_\alpha}^{\nu)\gamma} \\
  &\ - 2 {\mconn^{\beta}}_{\alpha\gamma} g^{\alpha(\mu}
  {\met{\dens{J}}_\beta}^{\nu)\gamma} + g^{\alpha\beta}
  {\mconn^\gamma}_{\beta\gamma} {\met{\dens{J}}_\alpha}^{\mu\nu} +
  g^{\alpha\beta} {\mconn^\gamma}_{\alpha\beta}
  {\met{\dens{J}}_\gamma}^{\mu\nu} \Bigr] \, ,
\end{split}
\end{equation}
where we use the formula
\[
{\mconn^\alpha}_{\beta\gamma} :=
\tfrac{1}{2} g^{\alpha\delta} \left( g_{\delta\beta,\gamma} +
  g_{\delta\gamma,\beta} - g_{\beta\gamma,\delta} \right)
\]
and the identity $0 = \metnabla_\lambda g_{\rho\sigma} =
g_{\rho\sigma,\lambda} - {\mconn^\tau}_{\rho\gamma} g_{\tau\sigma}
- {\mconn^\tau}_{\sigma\gamma} g_{\rho\tau}$.  Inserting~(\ref{eq:term1})
and~(\ref{eq:term2}) into~(\ref{eq:Einstein-equations-2})
and combining the partial derivatives
$\partial \met{\dens{J}}$ together with the products ``$\met{\dens{J}} \cdot
 \mconn$'' into the covariant derivatives $\metnabla
\met{\dens{J}}$, we get~(\ref{eq:Einst3}).

\subsection{The proof of formula (\ref{eq:rozklad-Ricci})}
\label{sec:proof-formula}

The symmetric part of the Ricci tensor of the affine connection $\Gamma$ is
given by:
\begin{equation}
  \label{eq:Ricci}
  K_{\mu\nu} = \partial_\rho {\Gamma^\rho_{\mu\nu} - \partial_{(\mu}}
    \Gamma^\rho_{\nu)\rho} +
    {\Gamma^\rho}_{\sigma\rho}{\Gamma^\sigma}_{\mu\nu} -
    {\Gamma^\rho}_{\sigma\mu}{\Gamma^\sigma}_{\rho\nu} \ .
\end{equation}
Inserting the decomposition $\Gamma = \mconn + f$, we get
\[
\begin{split}
  K_{\mu\nu} &= \stackrel{\met{K}_{\mu\nu}}{\overbrace{\partial_\rho
      {\mconn^\rho_{\mu\nu} - \partial_{(\mu}} \mconn^\rho_{\nu)\rho}
      + {\mconn^\rho}_{\sigma\rho}{\mconn^\sigma}_{\mu\nu} -
      {\mconn^\rho}_{\sigma\mu}{\mconn^\sigma}_{\rho\nu}}} \\
&\ + \stackrel{\met\nabla_\rho
    {f^\rho}_{\mu\nu}}{\overbrace{\partial_\rho {f^\rho}_{\mu\nu} +
      {\mconn^\rho}_{\rho\sigma}
      {f^\sigma}_{\mu\nu} - 2 {\mconn^\sigma}_{\rho(\mu} {f^\rho}_{\nu)\sigma}}} \\
  &\ - \stackrel{\met\nabla_{(\mu}
    {f^\rho}_{\nu)\rho}}{\overbrace{\partial_{(\mu}
      {f^\rho}_{\nu)\rho} + {\mconn^\sigma}_{\mu\nu}
      {f^\rho}_{\rho\sigma}}} +
  {f^\rho}_{\sigma\rho}{f^\sigma}_{\mu\nu} -
  {f^\rho}_{\sigma\mu}{f^\sigma}_{\rho\nu} \\
  &= \met{K}_{\mu\nu} + \met\nabla_\rho {f^\rho}_{\mu\nu} -
  \met\nabla_{(\mu} {f^\rho}_{\nu)\rho} + {f^\rho}_{\sigma\rho}{f^\sigma}_{\mu\nu}
- {f^\rho}_{\sigma\mu}{f^\sigma}_{\rho\nu} \ ,
\end{split}
\]
which proves (\ref{eq:rozklad-Ricci}).

\subsection{The proof of formula (\ref{eq:nonmetricity})}
\label{sec:proof-nonmetricity}

We will show that $\Gamma = \mconn + f$, with the nonmetricity tensor
$f$ given by formula~(\ref{eq:f(J,g)}), is the only solution of the
nonmetricity equations~(\ref{eq:nonmetricity}). As was already
mentioned in Appendix~\ref{sec:affine-principle} (cf.~the
equations~\eqref{eq:EL-aff-2-cov-a} and~\eqref{eq:EL-aff-2-cov-b}), the
equations~(\ref{eq:nonmetricity}) are equivalent to
\begin{equation}
  \label{eq:dpi=Jf}
  \nabla_\lambda \pi^{\mu\nu} = {\dens{J}_\lambda}^{\mu\nu} - \tfrac{2}{3}
  \delta_\lambda^{(\mu} {\dens{J}_\rho}^{\nu)\rho} \ .
\end{equation}
Plugging $\Gamma = \mconn + f$ into these equations we get
\begin{equation}
\label{eq:covdev-decomp} \nabla_\lambda \pi^{\mu\nu} =
{f^{\mu}}_{\lambda\rho} \pi^{\nu\rho} +{f^{\nu}}_{\lambda\rho}
\pi^{\mu\rho} - {f^\rho}_{\lambda\rho} \pi^{\mu\nu} \ ,
\end{equation}
because the remaining terms combine to the covariant derivative
$\metnabla_\lambda \pi^{\mu\nu}$, which vanishes identically.

Contracting both sides with $\pi_{\mu\nu} = g_{\mu\rho} g_{\nu\sigma}
\pi^{\rho\sigma}$ gives
\begin{equation}
  \label{eq:covdev-contr}
  \pi_{\mu\nu} \nabla_\lambda \pi^{\mu\nu} = - 2 \left( 2k
    \sqrt{\abs{\det g}} \right)^2 {f^\rho}_{\lambda\rho} \ ,
\end{equation}
which allows us to rewrite~(\ref{eq:covdev-decomp}) as
\begin{equation}
  \label{eq:covdev-decomp2}
  {f^\mu}_{\lambda\rho} \pi^{\nu\rho} + {f^\nu}_{\lambda\rho}
  \pi^{\mu\rho} = \nabla_\lambda \pi^{\mu\nu} - \frac{\pi_{\rho\sigma} \pi^{\mu\nu}
  \nabla_\lambda \pi^{\rho\sigma}}{2 \left( 2 k
      \sqrt{\abs{\det g}} \right)^2} \ .
\end{equation}
Consider the following quantity:
\begin{equation}
  \label{eq:Ahelper-def}
  {A_\lambda}^{\mu\nu} := \frac{1}{2k \sqrt{\abs{\det \pi}}} \left(
      \nabla_\lambda \pi^{\mu\nu} - \frac{\pi_{\rho\sigma} \pi^{\mu\nu}
        \nabla_\lambda \pi^{\rho\sigma}}{2 \left( 2 k
          \sqrt{\abs{\det g}} \right)^2} \right) \ .
\end{equation}
Raising the lower index we obtain the following identities:
\begin{equation}
  \label{eq:A-od-f}
  \begin{split}
    A^{\lambda\mu\nu} &= f^{\mu\nu\lambda} + f^{\nu\mu\lambda} \\
    A^{\mu\lambda\nu} &= f^{\lambda\mu\nu} + f^{\nu\mu\lambda} \\
    - A^{\nu\mu\lambda} &= -f^{\mu\nu\lambda} - f^{\lambda\nu\mu} \ .
\end{split}
\end{equation}
Similarly as in the standard derivation of the formula for
the Christoffel symbols, we add them up. After lowering one index we
obtain:
\begin{equation}
  \label{eq:f-od-A}
  {f^\lambda}_{\mu\nu} = {A_{(\mu}}^{\lambda\rho} g_{\nu)\rho} -
  \frac{1}{2} g_{\mu\rho} g_{\nu\sigma} g^{\lambda\tau}
  {A_\tau}^{\rho\sigma} \ .
\end{equation}
Inserting~(\ref{eq:Ahelper-def}) into~(\ref{eq:f-od-A}), we get
\begin{equation}
  \label{eq:f-od-nabla-pi}
  \begin{split}
    {f^\lambda}_{\mu\nu} &= \frac{2k}{\sqrt{\abs{\det g}}} \Biggl[
    g_{\rho(\mu} \nabla_{\nu)} \pi^{\lambda\rho} - \frac{1}{2}
    \delta^\lambda_{(\mu} g_{\rho\sigma} \nabla_{\nu)}
    \pi^{\rho\sigma} \\
    &\ - \frac{1}{2} \left( g_{\mu\rho} g_{\nu\sigma}
      g^{\lambda\tau} \nabla_\tau \pi^{\rho\sigma} - \frac{1}{2}
      g_{\mu\nu} g_{\rho\sigma} g^{\lambda\tau} \nabla_\lambda
      \pi^{\rho\sigma} \right) \Biggr] \ .
  \end{split}
\end{equation}
Inserting~(\ref{eq:dpi=Jf}) into~(\ref{eq:f-od-nabla-pi})
gives~(\ref{eq:f(J,g)}), proving that it is the only solution
of~(\ref{eq:nonmetricity}).

\section{An example of the Legendre transformation}
\label{sec:transf-covect}

% \bfseries We have provided examples of various parts of transformation between
% $\lmata$ and $\laff$. Examples~\ref{sec:affine-lagr-vacu}
% and~\ref{sec:affine-vers-einst} show a total transformation, but for
% matter Lagrangians which do not contain connection coefficients.
% Example~\ref{sec:transf-from-lmata} and~\ref{sec:transf-from-laff}
% shows a transformation between $\lmata$ and $\laff$, which does not
% involve the transformation between $\mconn$ and $\Gamma$. \normalfont

We have provided an example of the Legendre transformation between $\lmatm$ and $\lmata$.

Consider the following metric matter Lagrangian $\lmatm$:
\begin{equation}
  \label{eq:lmatm-ex1}
  \begin{split}
    \lmatm(g,\mconn,\phi,\partial\phi) &= \sqrt{| \det g |}
    \metnabla_{(\mu} \phi_{\nu)} \metnabla^{(\mu} \phi^{\nu)} \\
    &= \sqrt{| \det g |}
    g^{\mu\lambda} g^{\nu\kappa} \metnabla_{(\mu} \phi_{\nu)}
    \metnabla_{(\lambda} \phi_{\kappa)} \ ,
  \end{split}
\end{equation}
where $\phi_\nu$ is a covariant matter field (using the anti-symmetrized
part of $\metnabla \phi$ leads to the Maxwell Lagrangian~(\ref{eq:E-M})
for the electromagnetic field $F_{\mu\nu} = \partial_\mu A_\nu -
\partial_\nu A_\mu$, see \cite{gr-gauge}).  We are going to carry out
the Legendre transformation between $\lmatm$ and $\lmata$.

The nonmetricity current $\met{\dens{J}}$ is equal to
\begin{equation}
  \label{eq:metJ-ex1}
  \begin{split}
    {\met{\dens{J}}_\lambda}^{\mu\nu} &= \frac{\partial
      \metnabla_\alpha \phi_\beta}{\partial {\mconn^\lambda}_{\mu\nu}}
    \frac{\partial \lmatm}{\partial \metnabla_\alpha \phi_\beta} \\
    &= -2
    \sqrt{| \det g|} g^{\alpha\rho} g^{\beta\sigma} \metnabla_{(\rho}
    \phi_{\sigma)} \delta^{(\mu}_\alpha
    \delta^{\nu)}_\beta \phi_\lambda \\
    &= -2 \sqrt{| \det g|} g^{\mu\rho} g^{\nu\sigma} \metnabla_{(\rho}
    \phi_{\sigma)} \phi_\lambda \ .
  \end{split}
\end{equation}
The corresponding nonmetricity tensor $f$ is, according to the
formula~(\ref{eq:f(J,g)}), equal to
% \begin{equation}  % WERSJA SZEROKA
%   \label{eq:f-ex1}
%   \begin{split}
%     {f^\lambda}_{\mu\nu} &= - 2 k \Bigl[ g^{\rho\lambda} \phi_\mu
%     \metnabla_{(\nu} \phi_{\rho)} + g^{\rho\lambda} \phi_\nu
%     \metnabla_{(\mu} \phi_{\rho)} - g^{\rho\alpha} g^{\lambda\beta}
%     g_{\mu\nu} \phi_\rho \metnabla_{(\alpha} \phi_{\beta)}
%     - g^{\rho\lambda} \phi_\rho \metnabla_{(\mu} \phi_{\nu)} +
%     \tfrac{1}{2} g^{\alpha\beta} g^{\rho\lambda} g_{\mu\nu} \phi_\rho
%     \metnabla_{(\alpha} \phi_{\beta)} \\
%     &\ + \tfrac{1}{3}
%     \delta^\lambda_\mu g^{\rho\alpha} \phi_\rho \metnabla_{(\nu}
%     \phi_{\alpha)}
%     + \tfrac{1}{3} \delta^\lambda_\nu g^{\rho\alpha} \phi_\rho
%     \metnabla_{(\mu} \phi_{\alpha)} - \tfrac{1}{2}
%     \delta^\lambda_{\mu} \phi_{\nu} g^{\alpha\beta}
%     \metnabla_{(\alpha} \phi_{\beta)} - \tfrac{1}{2}
%     \delta^\lambda_{\nu} \phi_{\mu} g^{\alpha\beta}
%     \metnabla_{(\alpha} \phi_{\beta)} \Bigr] \ .
%   \end{split}
% \end{equation}
\begin{equation}
  \label{eq:f-ex1}
  \begin{split}
    {f^\lambda}_{\mu\nu} &= - 2 k \Bigl[ g^{\rho\lambda} \phi_\mu
    \metnabla_{(\nu} \phi_{\rho)} + g^{\rho\lambda} \phi_\nu
    \metnabla_{(\mu} \phi_{\rho)}
- g^{\rho\alpha} g^{\lambda\beta}
    g_{\mu\nu} \phi_\rho \metnabla_{(\alpha} \phi_{\beta)} \\
&\ - g^{\rho\lambda} \phi_\rho \metnabla_{(\mu} \phi_{\nu)} +
    \tfrac{1}{2} g^{\alpha\beta} g^{\rho\lambda} g_{\mu\nu} \phi_\rho
    \metnabla_{(\alpha} \phi_{\beta)}
     + \tfrac{1}{3}
    \delta^\lambda_\mu g^{\rho\alpha} \phi_\rho \metnabla_{(\nu}
    \phi_{\alpha)} \\
    &\ + \tfrac{1}{3} \delta^\lambda_\nu g^{\rho\alpha} \phi_\rho
    \metnabla_{(\mu} \phi_{\alpha)} - \tfrac{1}{2}
    \delta^\lambda_{\mu} \phi_{\nu} g^{\alpha\beta}
    \metnabla_{(\alpha} \phi_{\beta)} - \tfrac{1}{2}
    \delta^\lambda_{\nu} \phi_{\mu} g^{\alpha\beta}
    \metnabla_{(\alpha} \phi_{\beta)} \Bigr] \ .
  \end{split}
\end{equation}

To perform the Legendre transformation from $\lmatm$ to $\lmata$,
we must express $\mconn$ as a function of $\Gamma$ and the matter
field, according to~(\ref{eq:f-ex1}). This is an extremely
difficult task. However, all we need is to compute $\lmata$ and to
express the field configurations in terms of the connection $\Gamma$.
For this purpose we have to compute the quadratic combinations of
$f$'s which give the difference between $\lmatm$ and $\lmata$ and
to express $\metnabla \phi$ as a combination of $\nabla \phi$ (and
possibly other) terms. The affine covariant derivative of $\phi$
is equal to
\begin{equation}
  \label{eq:nabla-phi-ex1}
   \nabla_\mu \phi_\nu = \metnabla_\mu \phi_\nu  -
  {f^\lambda}_{\mu\nu} \phi_\lambda \ .
\end{equation}
Inserting~(\ref{eq:f-ex1}) into~(\ref{eq:nabla-phi-ex1}) gives
\begin{equation}
  \label{eq:nablaphi}
  \begin{split}
    \nabla_\mu \phi_\nu &= \metnabla_\mu \phi_\nu - 2 k \Bigl[
    \tfrac{4}{3} g^{\rho\sigma} \phi_\rho \phi_\mu \metnabla_{(\nu}
    \phi_{\sigma)}
+ \tfrac{4}{3}
    g^{\rho\sigma} \phi_\rho \phi_\nu \metnabla_{(\mu} \phi_{\sigma)} \\
&\ - \tfrac{1}{2} g^{\rho\sigma} g^{\tau\eta} g_{\mu\nu} \phi_\rho
    \phi_\tau \metnabla_{(\sigma} \phi_{\eta)}
- g^{\rho\sigma}
    \phi_\rho \phi_\sigma \metnabla_{(\mu} \phi_{\nu)}
    - g^{\rho
      \sigma} \phi_{\mu} \phi_{\nu} \metnabla_{(\rho} \phi_{\sigma)}
    \Bigr]\ .
  \end{split}
\end{equation}
Correspondingly, we express the relation $\metnabla\phi =
\metnabla\phi(g,\phi,\nabla\phi)$ as a sum (with unknown
coefficients) of all possible 2-covariant (with two lower indices and no higher indices)
expressions linear in $\nabla\phi$, which could be
constructed algebraically from $g$, $\phi$ and $\nabla\phi$:
\begin{equation}
  \label{eq:metnablaphi-ansatz}
  \begin{split}
    \metnabla_\mu \phi_\nu &= \alpha_1 \nabla_{(\mu} \phi_{\nu)} +
    \alpha_2 \nabla_{[\mu} \phi_{\nu]} + \beta g_{\mu\nu}
    + \gamma_1
    g^{\rho\sigma} \phi_\rho \phi_\mu \left( \nabla_{(\nu}
      \phi_{\sigma)} \right) \\
    &\ + \gamma_2 g^{\rho\sigma} \phi_\rho \phi_\mu \left( \nabla_{[\nu}
      \phi_{\sigma]} \right)
    + \delta_1 g^{\rho\sigma} \phi_\rho
    \phi_\nu \left( \nabla_{(\mu} \phi_{\sigma)} \right) + \delta_2
    g^{\rho\sigma} \phi_\rho
    \phi_\nu \left( \nabla_{[\mu} \phi_{\sigma]} \right) \\
    &\ + \epsilon \phi_\mu \phi_\nu \ ,
  \end{split}
\end{equation}
where lower Greek letters denote the scalar coefficients which do not
depend on $\nabla\phi$, with the exception of $\beta$ and
$\epsilon$, which are linear in $\nabla\phi$. We shall find their
values by plugging this {\em Ansatz} into~(\ref{eq:nablaphi}),
expecting that all terms other than $\nabla_\mu \phi_\nu$ cancel,
and solving the resulting system of linear equations. This system reads:
\begin{equation}
  \label{eq:ansatz-eqn-system}
  \begin{split}
    1 &= \left( 1 + 2 \psi \right) \alpha_1 \\
    1 &= \alpha_2 \\
    0 &= k \Omega' \alpha_1 + \left( 1 + 3 \psi \right) \beta + \psi
    \Omega' \gamma_1 + \psi \Omega' \delta_1 + \frac{\psi^2}{k} \epsilon \\
    0 &= - \tfrac{8}{3} k \alpha_1 + \left( 1 - \tfrac{1}{3} \psi
    \right) \gamma_1 - \tfrac{1}{3} \psi \delta_1 \\
    0 &= - \tfrac{8}{3} k \alpha_1 - \tfrac{1}{3} \psi \gamma_1 +
    \left( 1 - \tfrac{1}{3} \psi \right) \delta_1 \\
    0 &= \left( 1 - \tfrac{1}{3} \psi \right) \gamma_2 - \tfrac{1}{3}
    \psi \delta_2 \\
    0 &= - \tfrac{1}{3} \psi \gamma_2 + \left( 1 - \tfrac{1}{3} \psi
    \right) \delta_2 \\
    0 &= k \Theta' \alpha_1 - \tfrac{4}{3} k \beta - \tfrac{5}{3} k
    \Omega' \gamma_1 - \tfrac{5}{3} k \Omega' \delta_1 + \left( 1 -
      \tfrac{7}{3} \psi \right) \epsilon \ ,
  \end{split}
\end{equation}
where a scalar invariant $\psi$, independent of the connection, is
defined as:
\[
  \psi := k g^{\alpha\beta} \phi_\alpha \phi_\beta
\]
and
\[
\begin{split}
\Psi' &:= g^{\mu\rho} g^{\nu\sigma} \nabla_{(\mu} \phi_{\nu)}
\nabla_{(\rho} \phi_{\sigma)} \\
\Omega' &:= g^{\mu\rho} g^{\nu\sigma} \phi_\mu \phi_\nu
\nabla_{(\rho} \phi_{\sigma)} \\
\Pi' &:= g^{\mu\rho} g^{\nu\kappa} g^{\sigma\tau} \phi_\kappa
\phi_\tau \nabla_{(\mu} \phi_{\nu)} \nabla_{(\rho} \phi_{\sigma)} \\
\Theta' &:= g^{\mu\nu} \nabla_{(\mu} \phi_{\nu)}
\end{split}
\]
are the scalar invariants of the $\Gamma$ connection (their counterparts
depending on the metric connection are denoted in the same way, but
without a prime). The solution of the system reads:
\begin{equation}
  \label{eq:ansatz-solution}
  \begin{split}
    \alpha_1 &= \frac{1}{1 + 2 \psi} \\
    \alpha_2 &= 1 \\
    \beta &= \frac{3\Theta'{\psi }^2 - 3k\Omega' + 23k\psi \Omega'}
    {\left( 1 + 2\psi \right)
      \left( 3 + 2\psi  - 17{\psi }^2 \right) } \\
    \gamma_1 &= \delta_1 = - \frac{8k}{\left( 1 + 2 \psi \right)
      \left( 3 - 2 \psi \right)} \\
    \gamma_2 &= \delta_2 = 0 \\
    \epsilon &=
- \frac{9k\Theta' + 21k\Theta'\psi - 18k\Theta'{\psi
        }^2 + 92k^2\Omega' + 168k^2\psi \Omega'}{\left(3 - 2\psi
      \right) \left( 1 + 2\psi \right) \left( 3 + 2\psi - 17{\psi
          }^2 \right) } \ .
  \end{split}
\end{equation}
One should note that for certain configurations, some of the
coefficients diverge. It does not mean that our procedure is
ill-defined in this case, it merely means that we have some additional
constrainst imposed on $\nabla \phi$ (once again, we refer
to~\cite{framework}).

Inserting the {\em Ansatz}~(\ref{eq:metnablaphi-ansatz}), with the
coefficients given by~(\ref{eq:ansatz-solution}), into~(\ref{eq:lmatm-ex1})
gives
\begin{equation}
  \label{eq:lmatm-od-aff}
  \begin{split}
    \lmatm(g,\phi,\nabla\phi)
    &= \sqrt{ | \det g | } \Biggl[ \alpha_1^2 \Psi' + 4 \beta^2 +
    \frac{\psi^2}{k^2} \epsilon^2 + 2 \left( 2 \alpha_1 +
      \frac{\psi}{k} \gamma_1 \right) \gamma_1 \Pi' \\
    &\ + \left( 2 \alpha_1 \epsilon + 4 \beta \gamma_1 + 4
      \frac{\psi}{k} \gamma_1
      \epsilon \right) \Omega' + 2 \alpha_1 \beta \Theta'
    + 2\frac{\psi}{k} \gamma_1^2 \Omega'^2
    \Biggr] \ .
  \end{split}
\end{equation}

% We will show now that the affine matter Lagrangian is of the form:
% \begin{equation}
%   \label{eq:lmata-ans-pre}
%   \begin{split}
%     \lmata (g,\Gamma,\phi,\nabla\phi) &= \sqrt{|\det g|} \bigl(
%     C_{\Psi'} \Psi' + C_{\Pi'} \Pi' +
%     C_{\Theta'^2} \Theta'^2 \\
%     &\ + C_{\Omega'\Theta'} \Omega'\Theta' + C_{\Omega'^2} \Omega'^2
%     \bigr) \ ,
%   \end{split}
% \end{equation}
% where the $C$ coefficients depend only on the $\psi$ invariant.

Calculating the term $\pi^{\mu\nu} \left( {f^\rho}_{\mu\nu}
  {f^\sigma}_{\rho\sigma} - {f^\rho}_{\mu\sigma} {f^\sigma}_{\nu\rho}
\right)$ (subtracted from $\lmatm$ in order to get $\lmata$) in terms
of $\metnabla \phi$ and substituting~(\ref{eq:metnablaphi-ansatz})
gives, after subtracting it from the density~(\ref{eq:lmatm-od-aff}) and
substituting~(\ref{eq:ansatz-solution}), the final result:
\begin{equation}
  \label{eq:lmata-vector-example}
  \begin{split}
    \lmata (g,\phi,\nabla\phi) &= \sqrt{|\det g|} \bigl( C_{\Psi'}
    \Psi'
    + C_{\Pi'} \Pi' + C_{\Theta'^2} \Theta'^2 \\
    &\ + C_{\Omega'\Theta'} \Omega' \Theta' + C_{\Omega'^2} \Omega'^2
    \bigr) \ ,
  \end{split}
\end{equation}
where
\begin{align*}
  C_{\Psi'} &= \frac{1 - 2 k \psi}{1 + 2 \psi} \ , \qquad
  C_{\Pi'} = \frac{ 64 k \psi^2 + 256 k^3 \psi^2 + 64 k^2 \psi \left( 3 + 2
      \psi \right) - 144}{9 + 12 \psi - 12
    \psi^2 } \ , \\
  C_{\Theta'^2} &=
  \begin{aligned}[t]
    &\frac{3 k^3 \psi^3 \left( 1 + 3 \psi \right)^2 + 6 \psi^2 \left(
        3 + 2 \psi - 11 \psi^2 \right)}{\left( 1 + 2 \psi
      \right)^2 \left( 3 + 2 \psi - 17 \psi^2 \right)} \\
&+ \frac{3 k^2 \psi^2 \left( 9 + 40
        \psi + 21 \psi^2 - 54 \psi^3 \right)}{\left( 1 + 2 \psi
      \right)^2 \left( 3 + 2 \psi - 17 \psi^2 \right)} \\
    &+ \frac{\psi k \left( 9 + 12 \psi - 116 \psi^2 - 122 \psi^3 + 256
        \psi^4 \right) }{\left( 1 + 2 \psi \right)^2 \left( 3 + 2 \psi
        - 17 \psi^2 \right)} \ ,
  \end{aligned} \\
  C_{\Omega'\Theta'} &=
  \begin{aligned}[t]
    &\frac{ 8 \psi \left( k + 4 k^2 \right) - 12 k }{\left( 3 - 2
        \psi \right) \left( 1 + 2 \psi \right)^2} \\
    &+ \frac{108k - 18 k \psi \left( 14 + 15 k \right) + 12 k \psi^2
      \left( 1 + 46 k + 88 k^2 \right)}{(3 - 2 \psi) \left( 1 +
        2 \psi \right)^2 \left( 3 + 2 \psi - 17 \psi^2 \right)^2} \\
&+ \frac{2 k \psi^3 \left( 170 + 665 k
        - 2580 k^2 - 92 k^3 \right)}{(3 - 2 \psi) \left( 1 +
        2 \psi \right)^2 \left( 3 + 2 \psi - 17 \psi^2 \right)^2} \\
    &- \frac{ 2 k \psi^4 \left( 1180 - 2358 k + 2276 k^2 + 444 k^3
      \right)}{(3 - 2 \psi) \left( 1 + 2 \psi \right)^2 \left( 3 +
        2 \psi - 17 \psi^2 \right)^2} \\
&- \frac{16 k \psi^5 \left(34 + 277 k + 429 k^2 - 63 k^3
      \right) }{(3 - 2 \psi) \left( 1 + 2 \psi \right)^2 \left( 3 +
        2 \psi - 17 \psi^2 \right)^2} \ ,
  \end{aligned} \\
  C_{\Theta'^2} &=
  \begin{aligned}[t]
    &\frac{1620 k^2 - 216 k^2 \psi \left( 243 - 80 k \right) + 12 k^2
      \psi^2 \left( 6115 + 7118 k + 1748 k^2 \right)}{3 \left( 3 - 2
        \psi \right)^2 \left( 1 + 2 \psi \right)^2 \left( 3
        + 2 \psi - 17 \psi^2 \right)^2} \\
    &+ \frac{8 k^2 \psi^3 \left( 16300 - 17125 k + 25120 k^2 + 1058
        k^3 \right)}{3 \left( 3 - 2 \psi \right)^2 \left( 1 + 2
        \psi
      \right)^2 \left( 3 + 2 \psi - 17 \psi^2 \right)^2} \\
&- \frac{16 k^2 \psi^4 \left( 7403 + 29604 k - 11204 k^2
        - 1932 k^3 \right)}{3 \left( 3 - 2 \psi \right)^2 \left( 1 + 2
        \psi
      \right)^2 \left( 3 + 2 \psi - 17 \psi^2 \right)^2} \\
    &+ \frac{96 k^2 \psi^5 \left( 646 + 1433 k - 2240 k^2 + 294 k^3
      \right)}{3 \left( 3 - 2 \psi \right)^2 \left( 1 + 2 \psi
      \right)^2 \left( 3 + 2 \psi - 17 \psi^2 \right)^2} \ .
  \end{aligned}
\end{align*}

The transformation of $\lmata$ to $\laff$ consists in solving a
nonlinear (because of the terms like $\psi^3$, $\psi^4$ and
$\psi^5$) problem.

\noindent\textbf{Remark:} The transformation between $\lmata$ and $\lmatm$
is practical only for the
Lagrangians quadratic in the covariant derivatives, since only then we
have a system of linear equations to solve. We also want to
mention that for such Lagrangians, our transformation is of higher
order. That is, taking $\nabla\phi$, $\metnabla\phi$ and
$\phi$ itself to be of the first order in $\phi$, we have
\begin{equation}
  \label{eq:higher-order}
  \nabla \phi - \metnabla \phi = O ( \phi^3 ) \ .
\end{equation}
The proof is easy: the nonmetricity tensor is proportional to the
derivative of $\nabla \phi \nabla \phi$ over $\Gamma$, hence
it is of the order $O(\phi^2)$. Thus, the difference $\nabla \phi
- \metnabla \phi = ``f\cdot\phi$'' is of the order $O(\phi^3)$.

%\bibliography{equivalence}
%\bibliographystyle{apsrev}

\end{document}